\documentclass[a4paper,11pt]{article}

\usepackage{jheppub} 

\usepackage[T1]{fontenc} 

\usepackage{hyperref}

\usepackage{amsmath}

\usepackage{extarrows}
\usepackage{subcaption}
\usepackage{slashed}
\usepackage{listings}
\usepackage[titletoc]{appendix}
\usepackage{float}
\usepackage{multirow}
\usepackage{mathrsfs}
\usepackage{graphicx}
\graphicspath{{./figs/}}
\usepackage[compat=1.0.0]{tikz-feynman}

\newcommand{\angl}[1]{\left(#1\right)}
\newcommand{\sangl}[1]{\left[#1\right]}
\newcommand{\bangl}[1]{\left\{#1\right\}}

\title{\boldmath Space-like Electromagnetic Form Factors of Lambda- and Sigma-Baryons from Quark-Diquark Faddeev Equations}

\author[a]{Langtian Liu}
\author[a,b]{Christian S. Fischer}

\affiliation[a]{Institut für Theoretische Physik, Justus-Liebig-Universität Gie{\ss}en, 35392 Gie{\ss}en, Germany}
\affiliation[b]{Helmholtz Forschungsakademie Hessen für FAIR (HFHF), GSI Helmholtzzentrum für Schwerionenforschung, Campus Gie{\ss}en, 35392 Gie{\ss}en, Germany}

\emailAdd{Langtian.Liu@physik.uni-giessen.de}
\emailAdd{christian.fischer@theo.physik.uni-giessen.de}

\definecolor{webred}{rgb}{0.75,0,0}

\abstract{
	An important goal of ongoing and future experiments is to explore spectra and transition form factors 
	of baryons with non-zero strangeness. Of particular interest
	is the transition form factor$\gamma^{(*)} \Sigma^0 \rightarrow \Lambda$ in the time-like momentum region
	that can be extracted from Dalitz decays. On the road towards a theoretical description of these form 
	factors we extend a covariant dynamical quark-diquark model for the baryon Faddeev equation to the 
	strange-quark sector. Based on an excellent description of the mass spectrum of selected baryon 
	octet and decuplet states and reasonable results for the nucleon form factors we determine the elastic 
	electromagnetic form factors of $\Lambda$ and $\Sigma^+, \Sigma^0, \Sigma^-$ hyperons in the space-like 
	region as well as the ones for the octet transition $\gamma^{(*)} \Sigma^0 \rightarrow \Lambda$. 
	We discuss qualitative and quantitative features of the diquark-quark picture and compare systematically 
	with previous results from a three-body	Faddeev approach and lattice data where available.}


\begin{document} 
	
	\maketitle
	\flushbottom
	
	\section{Introduction}
	
	The spectrum and the structure of baryons provide us with important opportunities to explore the strong interaction
	described by quantum chromodynamics (QCD). One of the most established probes of their internal structure are 
	electromagnetic form factors (EMFFs) that allow to extract important global properties like charge radii and 
	magnetic moments but also reveal detailed insights into the transition from perturbative to non-perturbative 
	physics via their momentum dependence. While the electromagnetic form factors of baryons with zero net strangeness 
	are very well explored, see e.g. \cite{Tiator2011a,Aznauryan2012a,Eichmann2016e} and references
	therein, single- and double-strange baryons received much less attention so far but are an important goal of ongoing and future experiments \cite{Dudek:2012vr,Adamczewski-Musch2021a}. 
	
    In general, it is very interesting to systematically compare properties and structure of baryons with different
    net strangeness. Taking all quantum fluctuations into account, the quark-gluon interaction becomes flavour-dependent 
    and we may be able to extract valuable knowledge about the different roles of up/down and strange quarks inside 
    baryons by comparing (transition) form factors and structure functions across the light and strange baryon multiplets. 
    
    On the experimental side, it is one of the primary goals of the HADES collaboration to extract information on the 
    production and electromagnetic decays of hyperons in a FAIR-phase-0 experiment \cite{Adamczewski-Musch2021a,Lalik:2019yag}. 
    Here, the off-shell kinematics promises interesting insights into the time-like behaviour of the $\Sigma$ to $\Lambda$
    transition. Theoretically, time-like form factors of hyperons have only been begun to be explored in recent years \cite{Ramalho:2012im,Haidenbauer:2016won,Yang:2017hao,Cao:2018kos,Yang:2019mzq,Ramalho:2019ues} with various approaches. 
    The space-like region on the other hand is accessible with lattice QCD
    \cite{CSSMAndQCDSF/UKQCDCollaborations2014b,CSSMAndQCDSF/UKQCDCollaborations2014c}, has already been explored with
    functional methods (three-body Faddeev equations) \cite{Sanchis-Alepuz2016a,Sanchis-Alepuz2018b} and can also be 
    discussed by chiral perturbation theory (ChPT) \cite{Kubis:2000aa}, vector meson dominance 
    models (VMD) \cite{Ramalho:2012ad,Li2021b,Yan:2023yff} and light cone sum rules (LCSR) \cite{Liu:2009mb}. Furthermore, dispersion 
    methods are very well suited to connect time-like and space-like physics in the small-$Q^2$ momentum region
    \cite{Granados:2017cib,Junker:2019vvy,Lin:2022baj,Lin:2022dyu}.
    
    As explained in Ref.~\cite{Dobbs2017a}, the different yield rates of $\Lambda$ and $\Sigma$ hyperons in the experiments 
    may indicate that internal diquark correlations in flavour space are also manifest in coordinate/momentum 
    space. This suggests that a description in terms of a diquark-quark Faddeev equation may be a reasonable approach.
    The Poincar$ \acute{\text{e}} $ covariant quark-diquark Faddeev equation can be derived from an exact equation for six-point
    functions using a separable expression for quark-quark T-matrices in terms of diquark propagators and Bethe-Salpeter 
    amplitudes \cite{Oettel1998b,Bloch1999b}. In this approach, the diquarks are fully dynamical and have 
    internal structure. The interaction with the third quark proceeds via attractive quark-exchange, which leads to the formation of 
    baryon bound states and resonances. The diagrammatic expression of the equation is shown in Fig.\ref{fig:faddeevmomentum} 
    and explained in more detail below. There are good arguments both from theory and experiment for the existence of diquark
    correlations inside baryons, see e.g. Ref.~\cite{Barabanov2021c} for a recent review.

    Over time, the quark-diquark Faddeev equation approach has been employed successfully in the study of baryons
    \cite{Barabanov2021c,Eichmann2016e}, including mass spectra of octet and decuplet baryons
    \cite{Hellstern1997b,Oettel1998b,Nicmorus2009b,Eichmann2016f,Eichmann2018a,Chen2018c,Chen2019d,Liu2022b,Liu2023b},
    electromagnetic and axial form factors
    \cite{Cloet2009b,Nicmorus2010b,Eichmann2011b,Wilson2012b,Eichmann2012d,Segovia2014d,Segovia2014e,Lu2019b,Chen2019e,
    	Raya2021b,Cheng2022a,Chen2022b} as well as parton distribution functions \cite{Kusaka1997b,Bednar2018b,Chang2022b}.
    Interestingly enough, when it comes to hyperon EMFFs no such investigation has been published so far to our knowledge.
    This work is intended as a starting point to fill this gap, since we believe that theoretical guidance from a variety 
    of methods is beneficial for the interpretation of future experimental efforts in particular in the strange quark 
    sector \cite{Adamczewski-Musch2021a}. In this work, we therefore extend a model approach to the quark-diquark Faddeev 
    equation including parametrizations of the strange quark resembling actual solutions of corresponding Dyson-Schwinger 
    equations. An introduction to the quark-diquark Faddeev equation formalism of $\Lambda$ and $\Sigma$ is given in 
    Sec.~\ref{sec:faddeev-eq}, whereas in Sec.~\ref{sec:q_and_dq} we discuss the necessary quark and diquark ingredients.
    We have solved the quark-diquark Faddeev equations for selected baryon octet and decuplet states and discuss our results
    in Sec.~\ref{sec:mass-spec}. The obtained mass spectrum is in very good agreement with the experimental values.
    In Sec.~\ref{sec:emff} we then discuss our results for the elastic EMFFs of $\Lambda$ and $\Sigma^+, \Sigma^0, \Sigma^-$ 
    hyperons as well as the electromagnetic transition form factors of $\gamma^{(*)}\Sigma^0 \rightarrow \Lambda$ in the space-like 
    momentum region and compare our results with that obtained from other theoretical methods as well as with experimental 
    values (if possible). We summarize in Sec. \ref{sec:summary}.
	
	Throughout this work, we assume isospin symmetry and perform our calculations in the Euclidean metric.
	
	\section{The quark-diquark Faddeev equation of $\Lambda$ and $\Sigma$ baryons}
	\label{sec:faddeev-eq}
	
	\begin{figure}[h]
		\centering
		\includegraphics[width=0.7\linewidth]{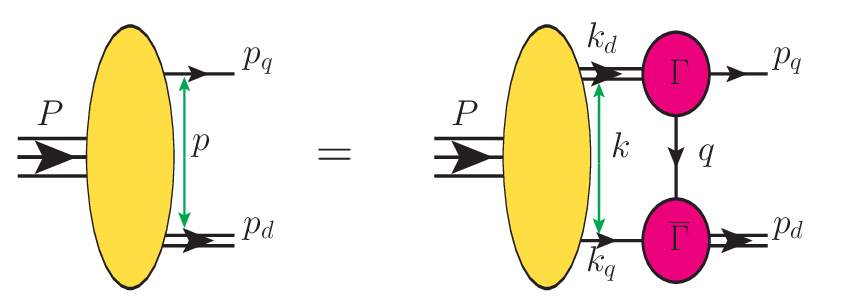}
		\caption{The Feynman diagram of the quark-diquark Faddeev equation. The meaning of each quantity is explained in the context. \label{fig:faddeevmomentum}}
	\end{figure}
	
	In the following we briefly recapitulate essential elements of the description of baryons using Faddeev equations. 
	Much more details as well as guidance for the numerical treatment of this type of equations can be found e.g., in 
	the review articles	\cite{Cloet:2013jya,Eichmann2016e,Sanchis-Alepuz:2017jjd} and references therein.
	
	Baryons obtain their quantum numbers from three valence quarks. Consequently the baryon wave function can be written 
	as a spinor of rank three $\psi_{\alpha \beta \gamma}$ ($\alpha, \beta, \gamma = 1, 2, 3, 4$ are the Dirac spinor indexes). 
	It is convenient to express this rank three spinor in terms of a quark-diquark form, which is a general three-fermion wave 
	function form \cite{A.B.Henriques1975b}.
	
	In this work, we focus on baryons with two light and one strange quark, i.e. the $\Lambda$ and $\Sigma$. Both have total 
	spin $J = \frac{1}{2}$ and their quark-diquark amplitude can be written as
	\begin{equation}
		\psi = \psi_1 + \psi_2 + \psi_3 \,.
	\end{equation}
	Here, the subscript indicates the bystander quark which does not join to form a diquark. In the case of a bystander quark 
	with index $i$ the amplitude is given by a summation of all allowed diquark channels:
	\begin{equation}
		\psi_i(p, P) = \sum_{a}^{} \sangl{\Gamma^{a}(l, p_d) D^{a}(p_d)} \sangl{\Phi_i^{a}(p,P) u(P)} \,, \quad i = 1, 2, 3 \,,
	\end{equation}
	where $\Phi_i^{a}(p, P)$ is a matrix valued function in Dirac space corresponding to the diquark $a$. It can be expanded 
	in an appropriate basis for different total spin and parity quantum numbers. Furthermore, $u(P)$ is the Dirac 
	spinor of a baryon with mass $M$ and satisfies the sum rule 
	$\frac{1}{2M} u(P) \bar{u}(P) = \frac{i M + \gamma \cdot P}{2 i M} \equiv \Lambda^+(P)$. 
	The momentum $P$ denotes the total momentum of the baryon and $p$ is the relative momentum between bystander quark 
	and diquark. The Bethe-Salpeter amplitude $\Gamma^{a}(l, p_d)$ of diquark $a$ depends on the total diquark momentum 
	$p_d$ and the relative momentum $l$ between two quarks inside the diquark. The quantity $D^a(p_d)$ is the diquark 
	propagator with momentum $p_d$.
	
	With given input, diquark propagators and Bethe-Salpeter amplitudes as well as quark propagators, the remaining
	unknown components in the baryon Bethe-Salpeter amplitude is determined by the quark-diquark Faddeev equation 
	in the form 
	\begin{equation}
		\Phi_i^{a} = \sum_{b}^{} \sum_{j \neq i}^{3} \int \dfrac{d^4k}{(2\pi)^4} K_{ij}^{ab}(k, p, P) G_j^{b}(k, P) \Phi_j^{b}(k, P) \,.
	\end{equation}
	Here 
	\begin{equation}
		G_j^{b}(k, P) = S_j(k_q) D^b(k_d) \,,
	\end{equation}
	is the combined quark and diquark propagator with quark momentum $k_q$ and diquark momentum $k_d$, whereas 
	\begin{equation}
		K_{ij}^{ab}(k, p, P) = \bar{\Gamma}^a_i(l_i, p_d) S^T(q) \Gamma^b_j(l_j, k_d) \,,
	\end{equation}
	is the quark exchange kernel. Here, $S^T(q)$ is the transposed quark propagator with momentum $q$. 
	The Feynman diagram of this quark-diquark Faddeev equation is displayed in Fig.\ref{fig:faddeevmomentum}.
	For $\Lambda$ and $\Sigma$ with spin-1/2 and positive parity, the decomposition of matrix valued functions for 
	scalar and axial-vector diquarks is shown in the appendix, Sec.\ref{subsec:basis}.

   An additional element implicit in the equations above is the flavour part of the baryon amplitude. In terms of
   quark and diquark components the $\Lambda$ and $\Sigma$ baryons are composed as follows:
	\begin{equation}\label{baryon-flavour}
		\begin{split}
			&\Lambda \sim \begin{bmatrix}
				\sangl{ud}s \\
				\frac{1}{\sqrt{2}}\angl{-\sangl{ds}u - \sangl{su}d} \\
				\frac{1}{\sqrt{2}}\angl{-\bangl{ds}u + \bangl{su}d}
			\end{bmatrix}\,, \quad
			\Sigma^+ \sim \begin{bmatrix}
				-\sangl{su}u \\
				\bangl{uu}s \\
				-\bangl{su}u
			\end{bmatrix} \,, \quad
			\Sigma^- \sim \begin{bmatrix}
				\sangl{ds}d \\
				\bangl{dd}s \\
				-\bangl{ds}d
			\end{bmatrix} \,, \\
			&\Sigma^0 \sim \begin{bmatrix}
				\frac{1}{\sqrt{2}}\angl{\sangl{ds}u - \sangl{su}d} \\
				\bangl{ud}s \\
				\frac{1}{\sqrt{2}} \angl{-\bangl{ds}u - \bangl{su}d}
			\end{bmatrix} \,, \quad
			\Xi^0 \sim \begin{bmatrix}
				-\sangl{su}s \\
				\bangl{su}s \\
				-\bangl{ss}u
			\end{bmatrix} \,, \quad
			\Xi^- \sim \begin{bmatrix}
				\sangl{ds}s \\
				\bangl{ds}s \\
				-\bangl{ss}d
			\end{bmatrix} \,,
		\end{split}
	\end{equation}
	where $\sangl{q_1 q_2} = (q_1 q_2 - q_2 q_1)/\sqrt{2}$ denotes a flavour-antisymmetric scalar and 
	$\bangl{q_1 q_2} = (q_1 q_2 + q_2 q_1)/\sqrt{2}$ a flavour symmetric axialvector diquark.

	\section{Quarks and Diquarks}
	\label{sec:q_and_dq}
	
	\begin{table}[h]
		\centering
		\begin{tabular}{c|c|c|c|c|c|c}
			\hline\hline
			$ \lambda_{u,d} $ & $ \bar{m}_{u,d} $ & $ b^{u,d}_0 $ & $ b^{u,d}_1 $ & $ b^{u,d}_2 $ & $ b^{u,d}_3 $ & $\varepsilon$ \\
			\hline
			0.566 & 0.00897 & 0.131 & 2.90 & 0.603 & 0.185 & 0.0001  \\
			\hline
			$ \lambda_{s} $ & $ \bar{m}_{s} $ & $ b^{s}_0 $ & $ b^{s}_1 $ & $ b^{s}_2 $ & $ b^{s}_3 $ & \\
			\hline
			0.817556 & 0.223 & 0.198323 & 1.19203 & 0.202049 & 1.19204 &  \\
			\hline\hline
		\end{tabular}
		\caption{The parameters of our representations of the $u,d$ and $s$ quark propagators, Eqs.\ref{q1}-\ref{q3}.\label{tab:q-para}}
	\end{table}

	In the past fifteen years, the quark-diquark Faddeev equation has been treated in several approaches with different 
	philosophies. Starting from a truncation of the quark-gluon interaction, one can solve the Dyson-Schwinger equation
	of the quark propagator in the complex momentum space, subsequently solve the corresponding Bethe-Salpeter equations
	for all diquarks needed in a given truncation and then use this input in the quark-diquark Faddeev equation to determine
	the baryon amplitudes. This program has been carried out with contact interactions between quarks as well as in 
	rainbow-ladder type truncations using models for the effective quark-gluon coupling, see e.g. 
	\cite{Eichmann:2007nn,Cloet:2013jya,Eichmann2016e} 
	for overviews. The second, and admittedly much more ad hoc method is to use model functions
	representing the most important properties of the quark and diquark propagators as well as their wave functions. 
	This type of modelling has been pioneered in \cite{Hellstern1997b,Oettel1998b} and is still in use today 
	(see e.g \cite{Cheng2022a,Chen2022b} and Refs. therein) due to its simplicity. 
	
	The present work is part of a research program that ultimately aims at calculating transition form factors in the 
	time-like region. Since simplicity is vital in first exploratory studies we resort to the diquark-quark model approach 
	and postpone more fundamental rainbow-ladder type calculations to future works.

	We therefore use parametrized quarks and diquarks. For $u,d$ quarks, we adopted the original forms and parameters which 
	have been shown to describe the nucleons and $ \Delta $-baryons well \cite{Cloet2009b,Segovia2014d,Chen2022b,Yin:2023kom}. 
	The quark propagator has the form
	\begin{align}\label{q1}
		S(p)=- i \slashed{p} \sigma_v(p^2) + \sigma_s(p^2) \,,\quad \sigma_v = \frac{\bar{\sigma_v}}{\lambda^2}\,, \quad \sigma_s = \frac{\bar{\sigma_s}}{\lambda} \,,
	\end{align}
	with propagator functions $\sigma_v$ and $\sigma_s$ and 
	\begin{align}
		&\bar{\sigma_v}(x) = \dfrac{1}{x + \bar{m}^2} \sangl{1 - \mathcal{F}\angl{2 (x + \bar{m}^2)}} \,, \quad \mathcal{F}(x) = \dfrac{1 - \exp[-x]}{x} \,,\label{q2}\\
		&\bar{\sigma_s}(x) = 2 \bar{m} \mathcal{F} \angl{2 (x + \bar{m}^2)} + \mathcal{F} (b_1 x) \mathcal{F} (b_3 x) \sangl{b_0 + b_2 \mathcal{F} (\epsilon x)} \,, \quad x = \dfrac{p^2}{\lambda^2}\,,\label{q3}
	\end{align}
	and the parameters are shown in Tab.\ref{tab:q-para}.
	
	For the $s$ quark, we assume a similar functional form but adapt the parameters by fitting to an explicit solution 
	of the strange quark Dyson-Schwinger equation obtained in the truncation of Ref.\cite{Bernhardt:2023ezo}. 
	Its parameters are also shown in Tab.\ref{tab:q-para}.
	
\begin{figure}[t]
		\centering
		\includegraphics[width=0.7\linewidth]{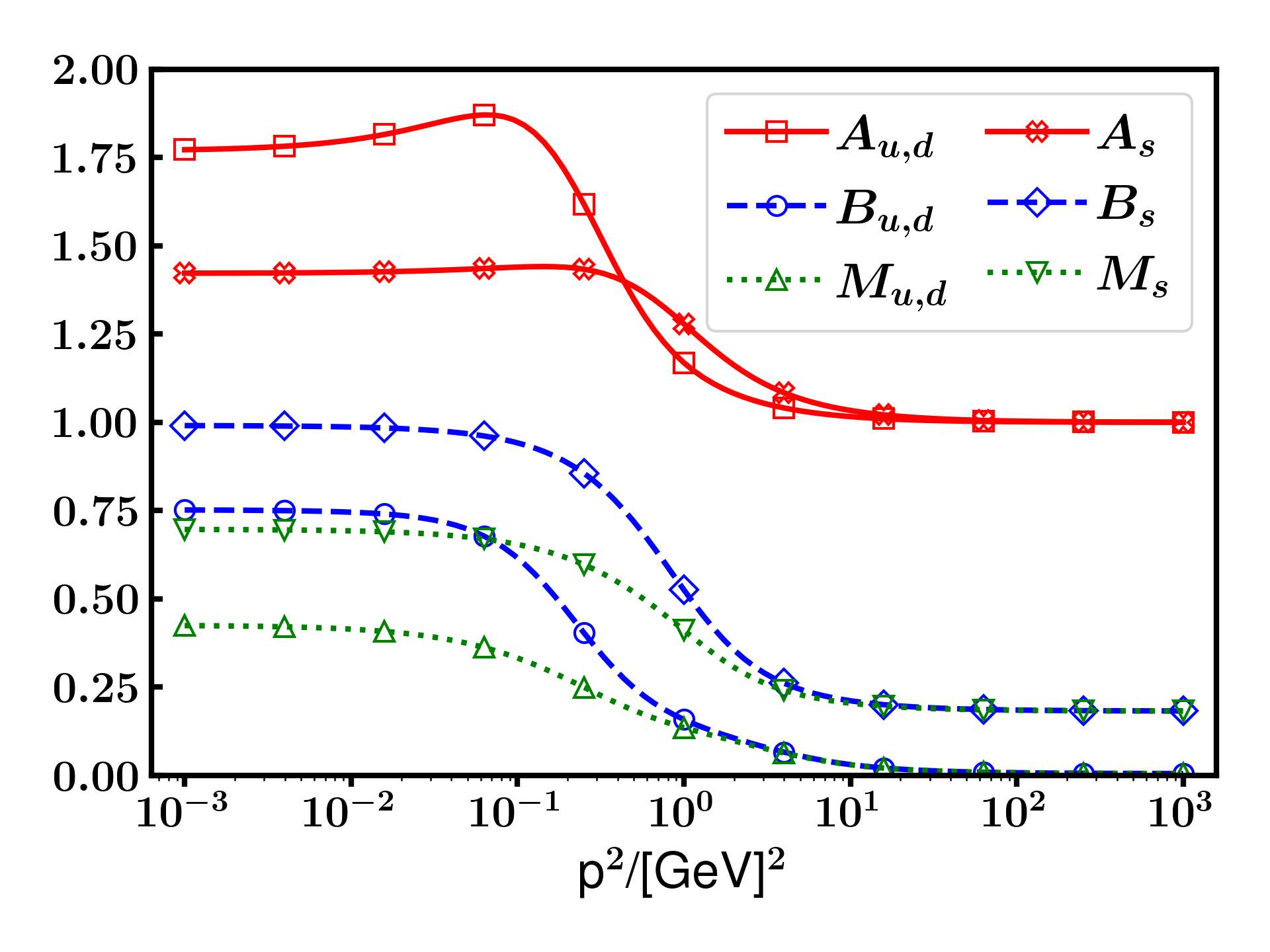}
		\caption{The dressing functions A (red, square), B (blue, circle) and M (green, upward triangle) of $u,d$ quarks and A (red, cross), B (blue, diamond) and M (green, downward triangle) $s$ quark. \label{fig:quark-func}}
	\end{figure}

	Defining the inverse quark propagator as 
	\begin{equation}
		S^{-1}(p) = i \slashed{p} A(p^2) + B(p^2) = A(p^2) \angl{i \slashed{p} + M(p^2)}\,,
	\end{equation}
	the dressing functions $A, B, M$ of $u,d$ and $s$ quarks are depicted in Fig.~\ref{fig:quark-func} as a function
	of space-like momentum $p^2$. One clearly distinguishes the large momentum behaviour of the dressing functions 
	('current quarks') from the small momentum region, where dynamical chiral symmetry breaking and the associated 
	quark mass generation sets in. At zero momenta one may read off 'constituent quark' masses of the order of 400 MeV 
	for the	up/down quarks and 700 MeV for the strange quark mass. 
		
	As for diquarks, the scalar ($0^+$) and axial vector ($1^+$) diquarks' propagators are parameterized as
	\begin{equation}
		D_{0^+}(k) = \dfrac{1}{M_{0^+}^2}\mathcal{F}(k^2/\omega_{0^+}^2)\,, \quad D_{1^+}^{\mu\nu}(k) = \left( g^{\mu\nu} + \dfrac{k^\mu k^\nu}{M_{1^+}^2}\right)\dfrac{1}{M_{1^+}^2}\mathcal{F}(k^2/\omega_{1^+}^2) \,,
	\end{equation}
	where $M_{0^+}, M_{1^+}$ and $\omega_{0^+},\omega_{1^+}$ are the masses and widths of scalar and axial vector diquarks 
	respectively. Since the diquarks originally emanate from T-matrices, their non-trivial on-shell behaviour is completely 
	carried by their Bethe-Salpeter wave functions and their propagators need to satisfy the condition 
	\begin{equation}
		\dfrac{d}{d k^2} \angl{D_{J^P}(k^2)}^{-1}\big |_{k^2 = 0} = 1 \quad \Rightarrow \quad \omega^2_{J^P} = \dfrac{1}{2} m^2_{J^P}
	\end{equation}
    for a free particle, see e.g. the appendix of Ref.~\cite{Segovia2014d}.
	
	The diquark amplitudes are parameterized as
	\begin{equation} \label{eq:dq-amp}
		\Gamma^a_{0^+}(l,p_d) = i g_{0^+} \gamma_5 C \mathcal{F}(l^2) \lambda^a \,, \quad \Gamma^{a,\mu}_{1^+}(l,p_d) = i g_{1^+} \gamma^\mu C \mathcal{F}(l^2) \lambda^a \,,
	\end{equation}
	where $l,p_d$ are the relative momentum between two quarks in diquark and total momentum of diquark respectively.
	$C$ is the charge conjugation operator.
	$\lambda^a$ is the diquark flavor wave function, which is shown explicitly in the appendix, Sec.\ref{subsec:dq-flavor}.
	$g_{0^+, 1^+}$ are the normalization constants of scalar and axial vector diquarks respectively. The diquark normalization condition is given by (Tr means trace over Dirac space)
	\begin{equation}\label{diquark-norm}
		\begin{split}
			N^2 &= \dfrac{1}{2}\dfrac{K^\mu}{2K^2}\dfrac{\partial}{\partial K^\mu}\dfrac{N_C^{dq}}{N_J}\text{Tr}\int \dfrac{d^4l}{(2\pi)^4} \left[ \bar{\Gamma}(l,-K_i) S(l + \eta_+K) \Gamma(l,K_i) S^T(-l + \eta_-K) \right] _{K=K_i}\,,
		\end{split}
	\end{equation}
	where $ N_C^{dq} = 1 $ is the normalized color factor of the diquark and $ N_J = 2J+1 $ counts its degrees of freedom.
	For the momentum partition parameters of quarks in the diquark we choose $ \eta_+ + \eta_- = 1 $ and $K_i$ is the on-shell 
	total momentum of diquark $K_i^2 = -M_{J^P}^2$. With the parameters in Tab.\ref{tab:dq-para}, we have explicitly 
	calculated the normalization constants of each diquark and listed the results in Tab.\ref{tab:dq-para}. These constants 
	also determine the coupling strength of each diquark to the two quark lines in the kernel of the quark-diquark 
	Faddeev equation (see above).
	
	The only remaining parameters that are up to now undetermined are the masses of the (flavour anti-symmetric) scalar diquarks
	$\sangl{ud}, \sangl{ds}, \sangl{su}$ and the (flavour-symmetric) axial vector diquarks 
	$\bangl{uu}, \bangl{ud}, \bangl{dd}, \bangl{ds}, \bangl{su}, \bangl{ss}$. Since we
	work in the isospin symmetric limit these amount to five different mass parameters. These have to be fixed using
	experimental data for baryon masses. This is discussed in the next section.

	\begin{table}[t]
		\centering
		\begin{tabular}{c|c|c|c|c}
			\hline\hline
			$ M_{\sangl{ud}} $ & $ M_{\bangl{uu},\bangl{ud},\bangl{dd}} $ & $M_{\sangl{ds}, \sangl{su}}$ & $M_{\bangl{ds},\bangl{su}}$ & $M_{\bangl{ss}}$\\
			\hline
			0.667 & 0.868 & 0.950 & 1.066 & 1.254 \\
			\hline\hline
			$ g_{\sangl{ud}} $ & $ g_{\bangl{uu},\bangl{ud},\bangl{dd}} $ & $g_{\sangl{ds}, \sangl{su}}$ & $g_{\bangl{ds},\bangl{su}}$ & $g_{\bangl{ss}}$ \\
			\hline
			21.52 & 15.03 & 17.64 & 15.97 & 15.62 \\
			\hline\hline
		\end{tabular}
		\caption{Masses (in GeV) and dimensionless coupling constants of diquarks. Whereas the latter are determined entirely 
			by the normalisation condition, Eq.~(\ref{diquark-norm}), the former are fixed by matching the masses of selected 
			baryon octet and decuplet states to experimental data. See main text for details. \label{tab:dq-para}}
	\centering\vspace*{5mm}
	\begin{tabular}{c|c|c|c|c|c|c|c}
		\hline\hline
		& $N(940)$ & $\Delta(1232)$ & $\Lambda(1116)$ & $\Sigma(1193)$ & $\Sigma(1385)$ & $\Xi(1315)$ & $\Omega(1672)$ \\
		\hline
		mass & 0.939(74) & 1.210(76) & 1.116(90) & 1.193(90) & 1.377(87) & 1.301(107) & 1.672(111) \\
		\hline\hline
	\end{tabular}
	\caption{Masses (in GeV) of selected baryon octet and decuplet states calculated from our quark-diquark Faddeev equation. 
		     The errors bars represent a measure for the model uncertainty generated from the variation of our input parameters,
		     the diquark masses, by 5\%. \label{tab:mass-spectrum}}
	\end{table}

	\section{Mass Spectrum}
	\label{sec:mass-spec}
	
	\begin{figure}[t]
	\centering
	\includegraphics[width=0.72\linewidth]{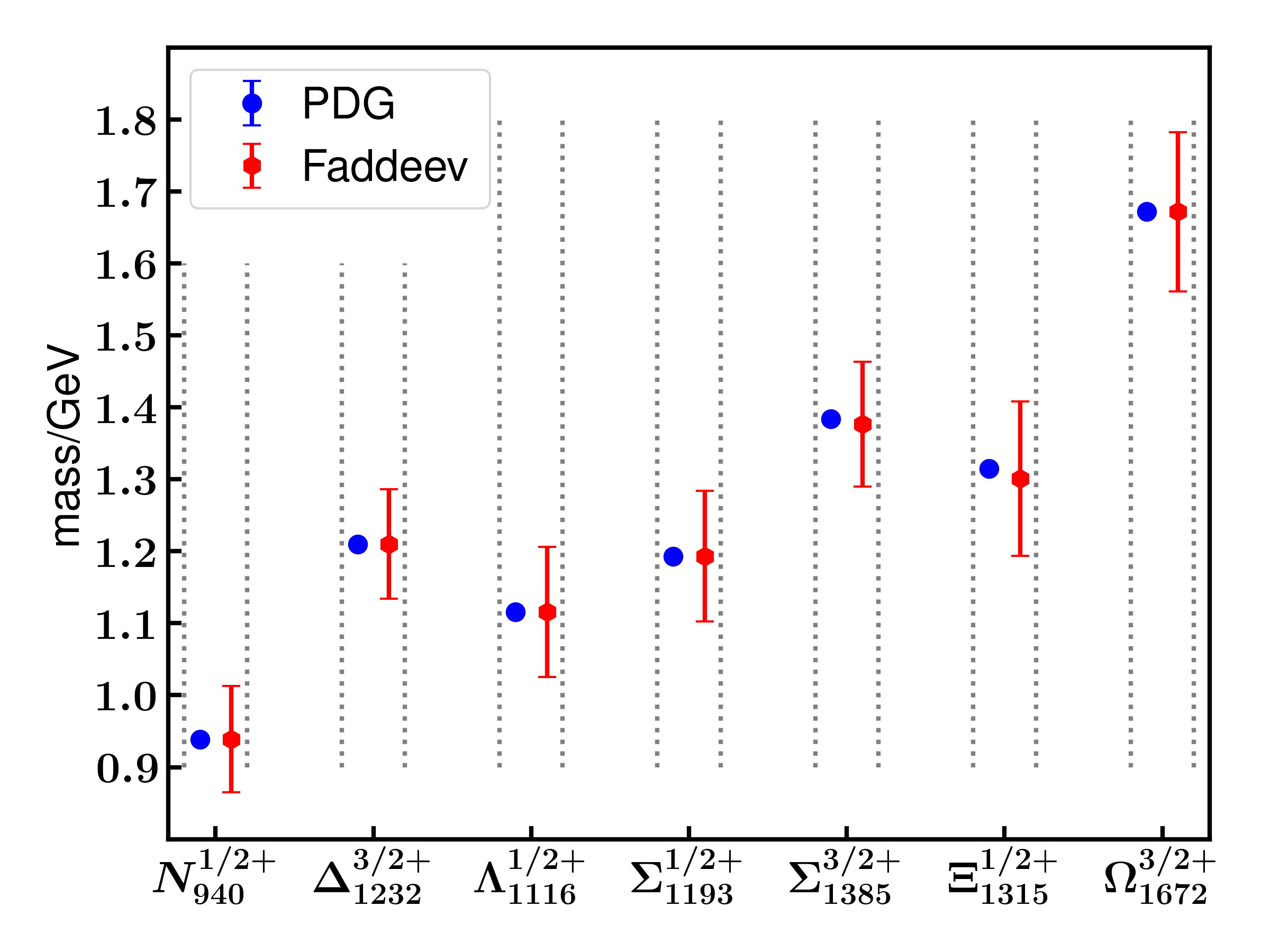}
	\caption{The mass spectrum of selected baryons calculated from our quark-diquark Faddeev equation (red hexagon) 
	and their comparison with the experimental values (blue circle). Error bars in our results reflect changes induced by variations 
		of the diquark mass	parameters by 5\%.}
	\label{fig:massspectrum}
    \end{figure}

	Using the parameters introduced in the Sec.\ref{sec:q_and_dq}, we solved the Faddeev equation of $N(940)$, $\Delta(1232)$, $\Lambda(1116)$, $\Sigma(1193)$, $\Sigma(1382)$, $\Xi(1315)$, $\Omega(1672)$ and obtained their mass spectrum. To this end,
	we have adapted the five different diquark masses listed in Tab.\ref{tab:dq-para} to have a best possible fit for the masses 
	of the five states $N(940)$, $\Delta(1232)$, $\Lambda(1116)$, $\Sigma(1193)$, $\Omega(1672)$. We then vary the best-fit 
	diquark	masses by 5\% in order to obtain an (obviously naive but still useful) measure for the systematic error of the 
	model. The results are listed in Tab.\ref{tab:mass-spectrum} and are compared to the experimental values in Fig.\ref{fig:massspectrum}. 
	
	From Fig.\ref{fig:massspectrum}, we find the predicted masses of $\Sigma(1385)$ and $\Xi(1315)$ from our quark-diquark 
	Faddeev equation approach in good agreement with their experimental values. As can be expected, we also find that the
	baryon masses are quite sensitive to the diquark masses. The 5\% variations in the diquark masses lead to variations in the
	baryon masses in the range of 6-8 \%. Fortunately, as we will see in the next sections, the electromagnetic form factors 
	of $\Lambda$ and $\Sigma$ will be much less sensitive to these variations.
	
	Finally, we would like to point out a particular appealing property of the diquark-quark model that has been noted 
	already in Ref.~\cite{Oettel1998b}: diquark correlations induce the mass split between $\Lambda$ and $\Sigma$ already on
	the level of model approaches such as the one pursued here and also on the level of rainbow-ladder type truncations  
	\cite{Eichmann2016e}. In genuine three-body treatments of the Faddeev equation, however, genuine rainbow-ladder type 
	interactions are flavour-blind and one has to go beyond rainbow-ladder to see any splitting
	\cite{Sanchis-Alepuz:2014wea,Sanchis-Alepuz2018b}. 	
	
	We therefore believe that the diquark-quark picture is an excellent starting point to go further and calculate the elastic 
	EMFFs of $\Lambda(1116)$, $\Sigma^+(1189)$, $\Sigma^0(1193)$, $\Sigma^-(1197)$ and the electromagnetic transition form 
	factors $\gamma^{(*)} \Sigma^0(1193) \rightarrow \Lambda(1116)$. The good agreement of the mass of the (decuplet) 
	$\Sigma(1385)$ with experiment suggests to go even further and also determine the electromagnetic transition form factors of
	$\gamma^{(*)} \Sigma(1385) \rightarrow \Lambda(1116)$. This is technically much more involved than the octet-octet transition
	and therefore left for future work. As mentioned in the introduction, in this work we furthermore restrict ourselves to the 
	analysis of the space-like momentum region. Ultimately, out goal is to extend this study to the technically more involved 
	time-like region in future work with applications on the analysis of data from HADES FAIR-phase-0 experiments in mind.

	\section{Electromagnetic form factors of $\Lambda$ and $\Sigma$}
	\label{sec:emff}
	
	In this section, we will discuss our results for the (space-like) EMFFs of $\Lambda-$ and $\Sigma-$type baryons. 
	This includes the elastic electromagnetic form factors of $\Lambda(1116)$, $\Sigma^+(1189)$, $\Sigma^0(1193)$, $\Sigma^-(1197)$ 
	and the corresponding transition form factors of the only allowed octet transition $\gamma^{(*)}\Sigma^0(1193) \rightarrow \Lambda(1116)$. 
	
	The electromagnetic currents between two spin-1/2 baryons with initial momentum $P_i$ and mass $M_i$ as well as final 
	momentum $P_f$ and mass $M_f$ can be written as ($e$ is the electric charge of positron)
	\begin{equation}\label{eq:em-current}
		\left\langle P_f \left| J^\mu \right| P_i \right\rangle = i e \bar{u}(P_f) \left( \gamma^\mu_T F_1(Q^2) + \dfrac{1}{M_f + M_i} \sigma^{\mu\nu}Q_\nu F_2(Q^2)\right) u(P_i)\,,
	\end{equation}
	where $Q = P_f - P_i$ is the photon momentum and $\gamma^\mu_T = \angl{g^{\mu\nu} - Q^\mu Q^\nu / Q^2} \gamma_\nu$.
	The Dirac, Pauli form factors $F_1, F_2$ and the Sachs electric, magnetic form factors $G_E, G_M$ are related via
	\begin{align}
		&G_E = F_1 - \dfrac{Q^2}{(M_f + M_i)^2} F_2 \,, \\
		&G_M = F_1 + F_2 \,. \label{magnetic}
	\end{align}
	
	Here we should comment on the unit of the magnetic form factor.	From the electromagnetic current, Eq.\eqref{eq:em-current},
	one reads of that the Pauli form factor $F_2$ is expressed in units of $e \hbar / (M_f + M_i)$ where $\hbar$ is the Plank constant.	However, conventionally, 
	one prefers to use common units for the magnetic form factors of all baryons and expresses the magnetic moment in units 
	of the nuclear magneton $e \hbar / 2 M_N$ where $M_N$ is the proton mass. In this work, we follow this convention and 
	also express the magnetic form factors of $\Lambda$, $\Sigma$ and $\gamma^{(*)} \Sigma^0(1193) \rightarrow \Lambda(1116)$ in units 
	of the nuclear magneton. Thus the magnetic form factors $G_M$ of Eq.~(\ref{magnetic}) are further multiplied by a 
	factor $2 M_N / (M_f + M_i)$.
	
	Technically, the procedure to couple an external field to a Bethe-Salpeter equation is called {\it gauging} of the equation 
	and was introduced in Refs.~\cite{Haberzettl:1997jg,Kvinikhidze:1998xn,Kvinikhidze:1999xp,Oettel2000e,Oettel2000f}. This
	procedure ensures gauge invariance, prevents over-counting of diagrams and leads to charge conservation. The diagrammatic 
	elements we need to calculate the EMFFs in this approach have been discussed in detail in many works, see e.g.
	\cite{Oettel2000e,Cloet:2013jya,Eichmann2016e} and are summarised in Appendix \ref{sec:emffs-ingredients} 
	for the convenience of the reader. These expressions also contain a number of further parameters that are fitted to 
	the nucleon's electromagnetic form factors. This is discussed in Appendix \ref{sec:emffs-ingredients}, where we also 
	display the corresponding results.

	\subsection{Elastic EMFFs of $\Lambda$ and $\Sigma$}
	
	Although we cannot observe the elastic EMFFs of $\Lambda$ and $\Sigma$ experimentally, they are still worth investigating 
	for two reasons. On the one hand, they express our knowledge of the electromagnetic properties of the $\Lambda$ and $\Sigma$.
	Systematic comparison with other electromagnetic form factors within the baryon multiplets might reveal important
	information of the flavour dependence of their internal dynamics. 	
	On the other hand, the elastic EMFFs are a convenient tool to normalize the Faddeev amplitudes of $\Lambda$ and $\Sigma^0$,
	which is in turn a prerequisite for the calculation of their transition form factors 
	$\gamma^{(*)} \Sigma(1193) \rightarrow \Lambda(1116)$.
	
\begin{figure}[t]
		\includegraphics[width=\linewidth]{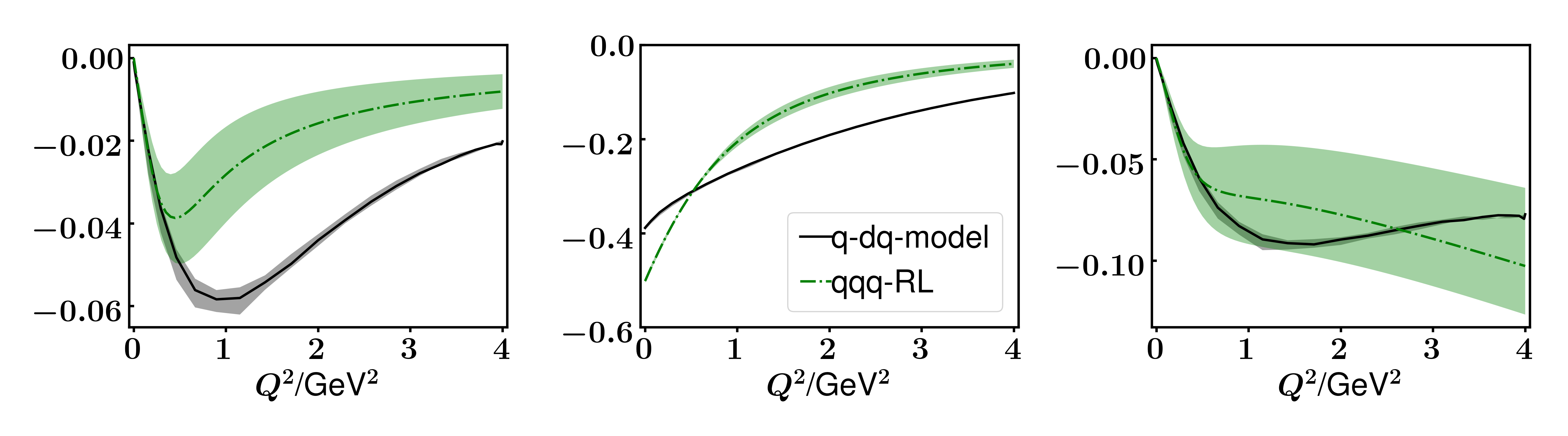}
		\caption{ The elastic electromagnetic form factors $G_E$ (left) and $G_M$ (center) and their ratio $G_E/G_M$ (right) 
		for the $\Lambda(1116)$. The magnetic form factor is expressed in units of the nuclear magneton. The black lines 
		are our results from the quark-diquark Faddeev equation ('q-dq-model'). The black band represents the error band 
		arising from a variation of diquark mass parameters by 5\%. The green dash-dotted lines and the corresponding error band 
		are results from a three-body Faddeev equation calculation ('qqq-RL') using an underlying effective running coupling 
		in rainbow-ladder (RL) approximation \cite{Sanchis-Alepuz2016a}. 
		\label{fig:emff-lambda}}
	\end{figure}

\begin{figure}[t]
		\includegraphics[width=\linewidth]{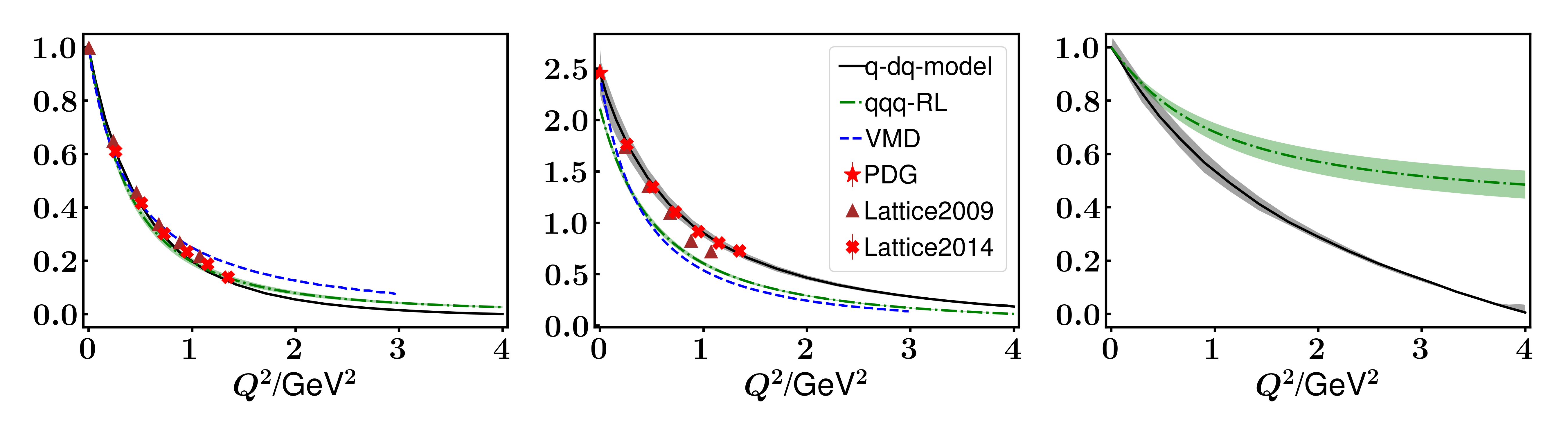}
		\\
		\includegraphics[width=\linewidth]{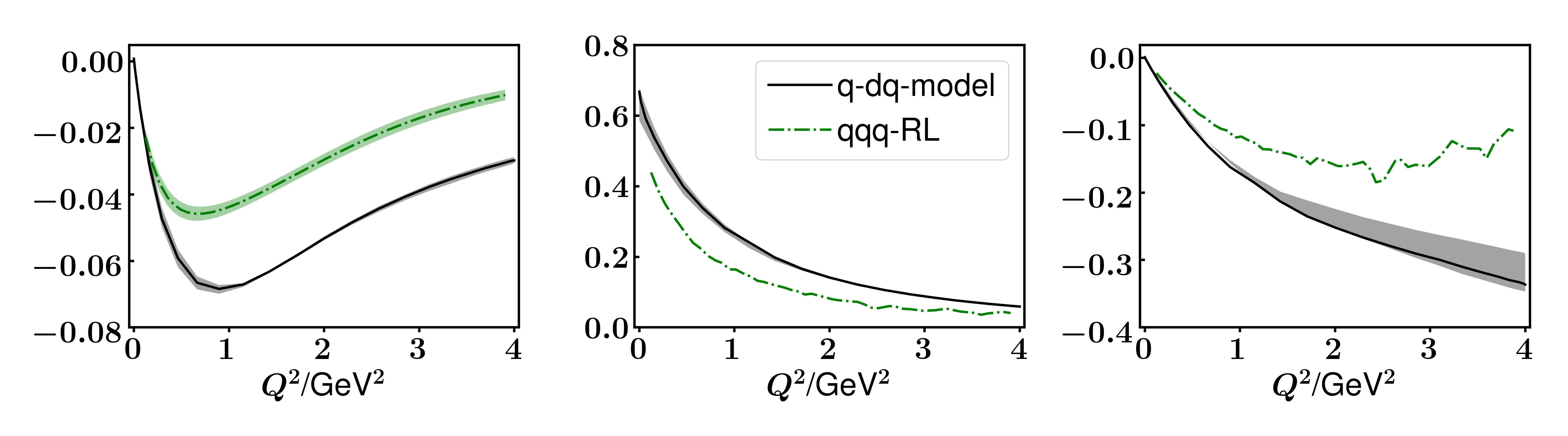}
		\\
		\includegraphics[width=\linewidth]{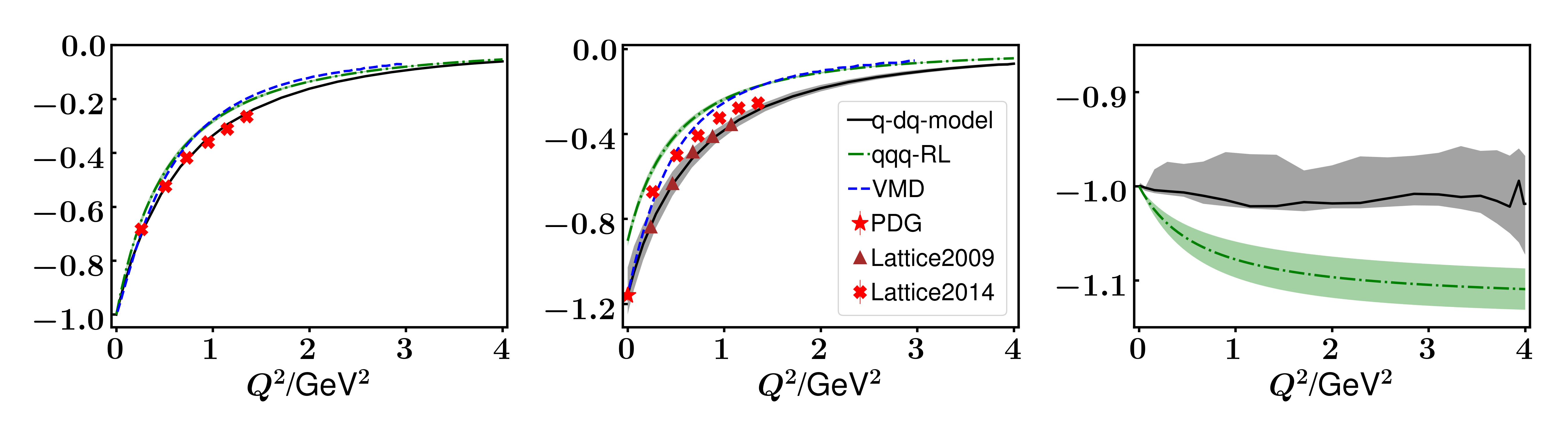}
		
		\caption{The elastic electromagnetic form factors $G_E$ (left), $G_M$ (center) and their ratio (right) of $\Sigma^+(1189)$ 
			(top), $\Sigma^0(1193)$ (middle) and $\Sigma^-(1197)$ (bottom). The magnetic form factor is expressed in units of 
			the nuclear magneton. The black lines are our results from the quark-diquark Faddeev equation ('q-dq-model'). 
			The black bands representan error estimate arising from the variation of diquark masses by 5\%. 
			The red star is the magnetic moment of $\Lambda^0(1116)$ from PDG \cite{pdg2022}. 
			Red crosses \cite{CSSMAndQCDSF/UKQCDCollaborations2014c,CSSMAndQCDSF/UKQCDCollaborations2014b} and brown 
			triangles \cite{Lin:2008mr} represent results from Lattice QCD calculations. The green dash-dotted lines and 
			bands ('qqq-RL') are again results from the three-body Faddeev equation \cite{Sanchis-Alepuz2016a}. The blue dashed lines are results 
			computed from a vector-meson dominance model \cite{Li2021b}. \label{fig:emff-sigma}
		}
	\end{figure}

	Since the $\Lambda$ and $\Sigma^0$ are charge zero, we normalize their Faddeev amplitudes from the flavor separated 
	electric form factors \cite{Cheng2022a}:
	\begin{equation}
		G_E(Q^2) = e_u G_E^u(Q^2) + e_d G_E^d(Q^2) + e_s G_E^s(Q^2) \,,
	\end{equation}
	where $e_u, e_d, e_s$ are the electric charges in units of $e$ and $G_E^u(Q^2), G_E^d(Q^2), G_E^s(Q^2)$ are the electric 
	charge form factors of $u, d, s$ quarks respectively. The normalization condition of $\Lambda$ and $\Sigma^0$ is then given by:
	\begin{equation}\label{eq:normalization-cond1}
		G_E^u(0) = G_E^d(0) = G_E^s(0) = 1\,.
	\end{equation}
	A convenient cross-check is provided by the fact that the normalization constant of $\Sigma^0$ calculated from
	Eq.~\eqref{eq:normalization-cond1} has to be the same as the normalization constants of $\Sigma^+$ and $\Sigma^-$ 
	calculated from the usual normalization conditions $G_E(0) = 1$ for $\Sigma^+$ and $G_E(0) = -1$ for $\Sigma^-$.
	This is indeed the case in our calculation. 
	The results for the elastic EMFFs of the $\Lambda$ and $\Sigma$ triplet are shown in Figs.~\ref{fig:emff-lambda} and 
	\ref{fig:emff-sigma} respectively. As with the baryon masses, our error bands arise from varying the diquark mass 
	parameters by 5\%.

	In Fig.\ref{fig:emff-lambda}, we compare the EMFFs of the $\Lambda(1116)$ from our quark-diquark Faddeev equation with 
	corresponding results from a three-body Faddeev equation \cite{Sanchis-Alepuz2016a}. 
	For the electric form factor, we 
	find almost identical results for small	momentum transfer $Q^2$. This also implies that the two methods predict the same 
	electric charge radius of the $\Lambda(1116)$. The large momentum behaviour, however, is similar in its functional form 
	but markedly different in size. A similar difference can be observed for the magnetic form factor. This difference in
	fall-off may be related to the different internal dynamics of the two approaches (i.e. the quark-gluon interaction
	explicit in the three-body approach and implicit in the quark-diquark model) but also may be generic to the differences
	between a three-body and a quark-diquark picture (or a combination of both). In our results for the corresponding nucleon
	EMFFs, discussed in Appendix Sec.~\ref{nucleon}, we do not see such large differences. Thus a definitive answer to this question
	may have to wait until a quark-diquark calculation using the same effective model as the three-body Faddeev equation is 
	available.\footnote{Note also that the present approach is not suitable (and not intended) to study the 
		behaviour of the form factors in the very large, perturbative momentum region, since our ansaetze for 
		the quark dressing functions do not include the proper logarithmic perturbative running. This is 
		different for more advanced diquark-quark approaches and the three-body Faddeev approach using an 
		underlying effective running coupling, see e.g. \cite{Eichmann2016e} for a detailed discussion on 
		the different approaches.}\\
	At small momenta the magnetic form factor $G_M$ of our quark-diquark Faddeev equation predicts a smaller decrease rate 
	than the three-body Faddeev equation method. This might be an effect due to diquark correlations decreasing the 
	size of the $\Lambda$ as compared to the three-quark picture. As a result, the magnetic form factor predicted by
	our quark-diquark Faddeev equation is harder than the three-body Faddeev equation prediction. Similar effects are seen
	in the nucleon, cf. our discussion in appendix \ref{nucleon}.

	In Fig.\ref{fig:emff-sigma}, we compare the EMFFs of $\Sigma^+$, $\Sigma^0$ and $\Sigma^-$ from our quark-diquark Faddeev 
	method with that from other theoretical methods such as the three-body Faddeev approach \cite{Sanchis-Alepuz2016a},
	Lattice QCD \cite{CSSMAndQCDSF/UKQCDCollaborations2014c,CSSMAndQCDSF/UKQCDCollaborations2014b,Lin:2008mr} and a 
	vector-meson dominance model \cite{Li2021b}. We also display the experimental values of the magnetic moment at 
	zero momentum \cite{pdg2022}. We obtain good agreement with the lattice calculations of $G_E$, $G_M$ for 
	$\Sigma^+$ and $\Sigma^-$. We interpret this as further indication that the diquark-quark picture is valid and our model
	provides for reasonable results. 

    An interesting difference can be seen in the ratios $\mu * G_E / G_M$ for the $\Sigma^{+,0}$ and $\Sigma^-$ baryons. 
	The ratio for $\Sigma^{+}$ decreases monotonically from one to zero in our quark-diquark Faddeev framework, i.e. the electric charge form factor $G_E$ 
	decreases much faster than the magnetic form factor $G_M$. 
	The three-body Faddeev equation framework \cite{Sanchis-Alepuz2016a} also exhibits a similar phenomenon, albeit with a less rapid decrease in the ratio. 
	In contrast, the ratio for $\Sigma^-$ starts at minus one and 
	stays constant within error bars in our quark-diquark Faddeev framework, i.e. the electric and magnetic form factors decreases with the same rate. 
	In the three-body Faddeev framework, the electric form factors even decrease slower than the magnetic form factors.
	Consequently, both the quark-diquark and three-body Faddeev frameworks display a larger decrease rate of electric form factors $G_E$ in $\Sigma^+$. 
	while they both show equal or smaller decrease rate of $G_E$ in $\Sigma^-$ comparing to the corresponding magnetic form factors $G_M$.
	A similar observation has been made in Ref.~\cite{Li2021b} in the time-like momentum region. A possible explanation for this 
	behaviour starts from the observation that the $\Sigma^+$ and $\Sigma^-$ have flavor wave functions of the same structure,
	see Eq.~(\ref{baryon-flavour}), but with the two $u$ quarks in $\Sigma^+$ replaced by two $d$ quarks in $\Sigma^-$. 
	The three possible flavour structures are dominated by the scalar $\sangl{su}$ diquark in the $\Sigma^+$ baryon and the 
	scalar $\sangl{ds}$ in $\Sigma^-$ \cite{Chen2019d}. Thus, approximately, the $\Sigma^+$ can be viewed as a two-body state 
	composed of a $u$ quark (with charge $+\frac{2e}{3}$) and a heavier $\sangl{su}$ diquark (with charge $+\frac{e}{3}$).
	The $\Sigma^-$ is well represented by a $d$ quark (with charge $-\frac{e}{3}$) and a heavier $\sangl{ds}$ diquark (with 
	charge $-\frac{2e}{3}$). Therefore, viewed from the rest frame of the respective diquarks, twice the amount of charge 
	is located in the center of the $\Sigma^-$ than in the center of the $\Sigma^+$ and the corresponding electric charge 
	form factor $G_E$ of $\Sigma^+$ will decrease more rapidly than that of the $\Sigma^-$, while the magnetic form factor 
	$G_M$ of $\Sigma^+$ will decrease slower than that of $\Sigma^-$. In combination, this leads to the different behaviors 
	of the ratio of EMFFs. The $\Sigma^0$ is in-between and therefore its form factor ratio falls off less rapidly than the
	one of the $\Sigma^+$. Similar qualitative features have been observed in lattice calculations \cite{Boinepalli:2006xd,CSSMAndQCDSF/UKQCDCollaborations2014c,CSSMAndQCDSF/UKQCDCollaborations2014b}.

	\begin{table}[t]
	
	\begin{tabular}{c||c|c|c|c|c||c|}
		& $\left\langle r_E^2 \right\rangle$ & $\left\langle r_E^2 \right\rangle_{\text{3b}}$ & $\left\langle r_E^2 \right\rangle_{\text{VMD}}$ & $\left\langle r_E^2 \right\rangle_{\text{ChPT}}$ & $\left\langle r_E^2 \right\rangle_{\text{disp}}$& $\left\langle r_E^2 \right\rangle_{\text{PDG}}$ \\\hline
		$\Lambda $  &0.036(14)	&0.04(1)  &0.012	&0.11(2)&-&-\\\hline
		$\Sigma^+$  &0.469(9)	&0.56(3)  &0.80(2)	&0.60(2)&-&-\\\hline
		$\Sigma^0$  &0.068(9)	&0.057(8) &0.10(1)	&-0.03(1)&-&-\\\hline
		$\Sigma^-$  &0.353(26)\phantom{ }	&0.45(3)  &0.70(2)	&0.67(3)\phantom{  }&-&0.61(15)\\\hline
	\end{tabular}\\\vspace*{5mm}

	\begin{tabular}{c||c|c|c|c|c|}
		& $\left\langle r_M^2 \right\rangle$ & $\left\langle r_M^2 \right\rangle_{\text{3b}}$ & $\left\langle r_M^2 \right\rangle_{\text{VMD}}$ 
		& $\left\langle r_M^2 \right\rangle_{\text{ChPT}}$& $\left\langle r_M^2 \right\rangle_{\text{disp}}$\\\hline
		$\Lambda $  &0.120(76)	&0.21(1) &0.18	&0.48(9) & 0.464(2)\\\hline
		$\Sigma^+$  &0.374(41)	&0.43(2) &-		&0.80(5) &-\\\hline
		$\Sigma^0$  &0.201(169)	&0.39(3) &-		&0.45(8) &-\\\hline
		$\Sigma^-$  &0.459(122)	&0.50(1) &-		&1.20(13)&-\\\hline
    \end{tabular}\\\vspace*{5mm}
	
	\begin{tabular}{c||c|c||c|}
		& $\mu$ 		& $\mu_{\text{3b}}$ & $\mu_{\text{PDG}}$\\\hline
		$\Lambda $ &-0.390(3)	&-0.435(5)			&-0.613(4)			\\\hline
		$\Sigma^+$ & 2.422(180)	& 1.82(2)			&2.458(10)			\\\hline
		$\Sigma^0$ & 0.630(48)	& 0.521(1)			& -					\\\hline
		$\Sigma^-$ &-1.145(106)	& -0.78(2)			&-1.160(25)			\\\hline
	\end{tabular}
	\caption{In the upper two tables we display our results for the mean-squared magnetic and electric charge radii expressed 
		in units of fm$^2$. We compare our results with the ones of the three-body approach \cite{Sanchis-Alepuz2016a}, available VMD results \cite{Yang:2019mzq,Yan:2023yff}, from chiral perturbation theory (ChPT) \cite{Kubis:2000aa} and dispersion theory (disp) \cite{Lin:2022baj} and 
		the only available experimental result for the $\Sigma^-$ \cite{pdg2022}. In the lower table we display our results 
		for the magnetic moments in units of the nuclear magneton and compare again with the three-body approach 
		\cite{Sanchis-Alepuz2016a} and experimental results\cite{pdg2022}. 
		\label{tab:charge-radii}}
\end{table}

	This qualitative discussion can be made more quantitative by determining the corresponding electric and magnetic charge radii. 
The mean-squared charge radii can be extracted from the form factors $G = G_{E,B}$ via
\begin{equation}
	\left\langle r^2 \right\rangle = -6 \dfrac{d}{d Q^2} \dfrac{G(Q^2)}{G(0)} \bigg | _{Q^2=0} \,.
\end{equation}
where the term $G(0)$ is set to one for form factors constraint to vanish at the origin. The obtained mean-squared charge radii of $\Lambda$ 
and $\Sigma$ from the quark-diquark equations are listed in Tab.\ref{tab:charge-radii} together with our results for the 
magnetic moments and their comparison with the corresponding results in the three body Faddeev approach \cite{Sanchis-Alepuz2016a}, 
available VMD results \cite{Yang:2019mzq,Yan:2023yff}, from chiral perturbation theory (ChPT) \cite{Kubis:2000aa} and dispersion theory (disp)
\cite{Lin:2022baj}and experimental values from PDG \cite{pdg2022}. 
Qualitatively, our results for the electric and magnetic charge radii agree very well with the ones from the three-body approach and the 
VMD-results, whereas the values from ChPT for most hyperons and the one from dispersion theory for the $\Lambda$ are larger. Taken at face
value this points towards possible meson cloud effects that are not accounted for in our approach and therefore result in somewhat too small 
charge radii.

It is nevertheless interesting to note that in both DSE/BSE approaches (and the VMD calculation) the 
electric and magnetic charge radii of the $\Lambda$ and the $\Sigma^0$ are quite different, although the quark content of both types of
baryons is the same. As already discussed in Ref.~\cite{Boinepalli:2006xd}, quark model explanations of this difference 
usually fall short (see also \cite{Sanchis-Alepuz2016a} for a discussion). Also different intermediate virtual transitions
between baryons (discussed in \cite{Boinepalli:2006xd}) cannot be the prime source of the difference in radii since both, 
the three-body Faddeev approach and our quark-diquark picture reproduce such a difference without baryonic intermediate 
states. Instead, our results give further support to the conclusion of Ref.~\cite{Sanchis-Alepuz2016a} that the difference
may be solely generated by the different symmetries in the flavour structure of the states. This needs to be explored further.

    An interesting difference between the three-body approach and our quark-diquark model concerns the magnetic moments. 
Here, the three-body results fall drastically short in comparison with the experimental results, whereas we could reproduce
those. The general explanation of the failure of the three-body approach is missing pion and kaon cloud effects. A pure
quark-core calculation has to fall short of the experimental magnetic moments, as convincingly argued in many works,
see e.g., \cite{Eichmann2011b,Eichmann:2011pv,Sanchis-Alepuz2016a} and references therein. So why is this different in the
quark-diquark model? It seems as if part of the effects of the pion and meson cloud, in particular static effects, can be
accommodated by the many parameters inherent of the model. Thus magnetic moments can be 'made' right by a suitable choice 
of parameters. Nevertheless, the explicit momentum dependent dynamics of the related diagrams in form factor calculations 
is not included in our model. Consequently, the smaller values of both approaches for the electric charge radius of the 
$\Sigma^-$ as compared to experiment (and ChPT) may be interpreted as due to missing meson cloud effects, as already discussed above. 

In order to make our numerical results for the electromagnetic form factors of $\Lambda$ and $\Sigma$ available for 
comparisons and also for further calculations, we provide corresponding fits. To this end we use similar fit functions 
as in Ref.~\cite{Sanchis-Alepuz2016a}: 
\begin{equation}\label{eq:fit-form}
	G(Q^2) = \dfrac{n_0 + n_1 Q^2}{1 + d_1 Q^2 + d_2 Q^4 + d_3 Q^6}\,.
\end{equation}
The best-fitted parameters for $\Lambda$ and $\Sigma$ are given in Tab.~\ref{tab:fit-params} in appendix \ref{app:fits}.
		
	\subsection{Transition form factors of the process $\gamma^{(*)} \Sigma^0 \rightarrow \Lambda$}
	
\begin{figure}[t]
		\begin{center}
		\includegraphics[width=0.50\linewidth]{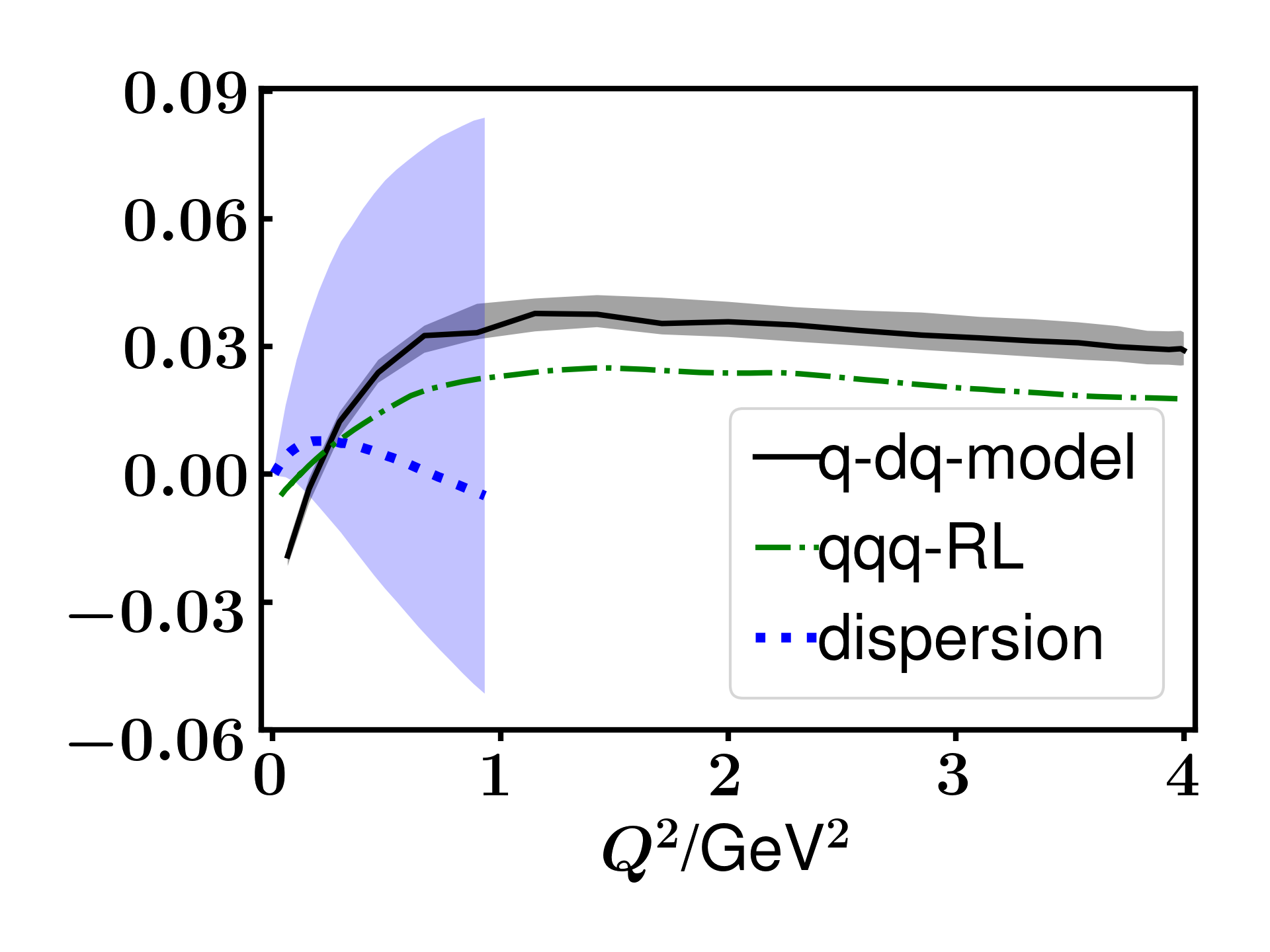}\hfill
		\includegraphics[width=0.50\linewidth]{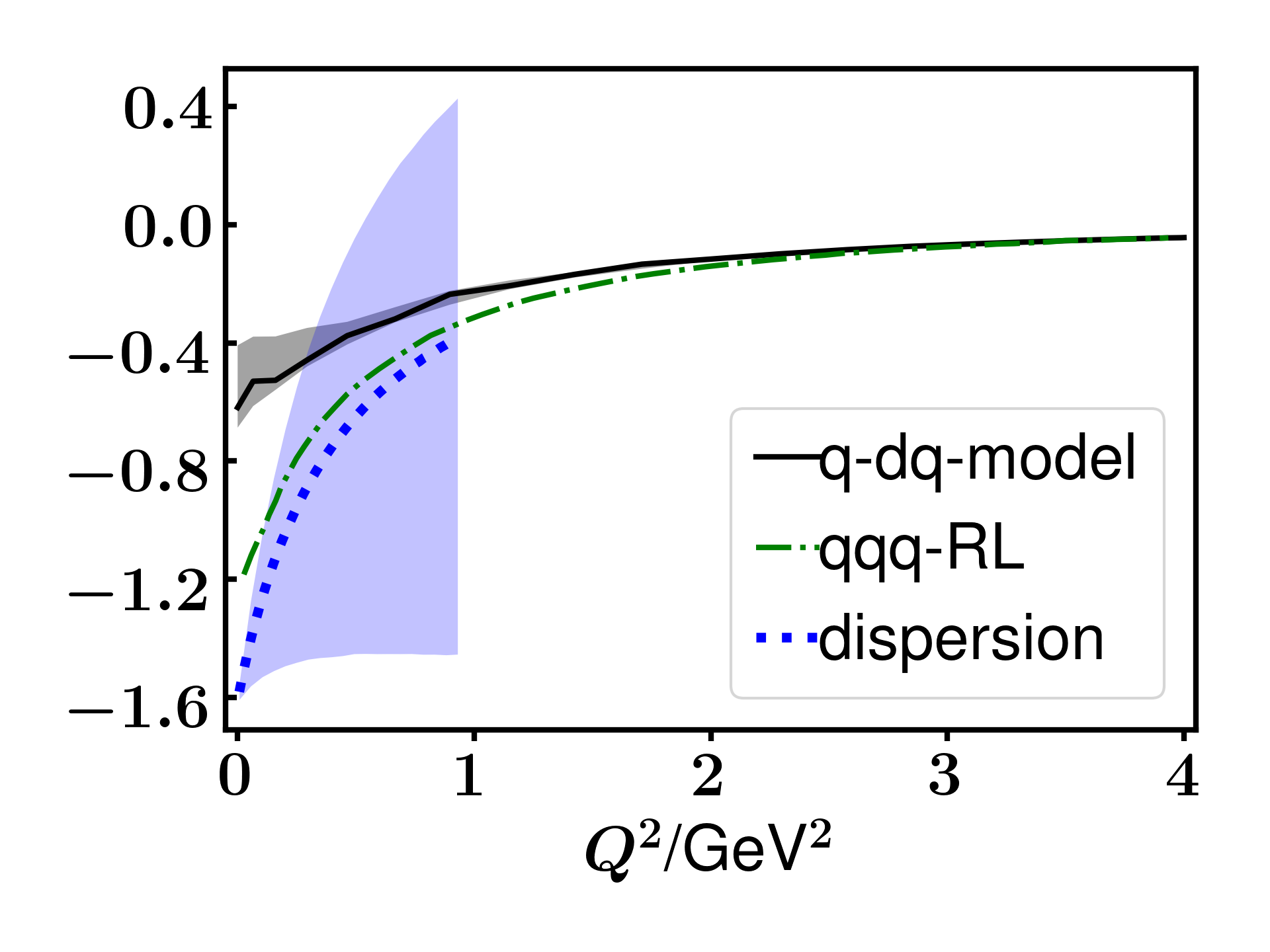}\hfill
		\includegraphics[width=0.50\linewidth]{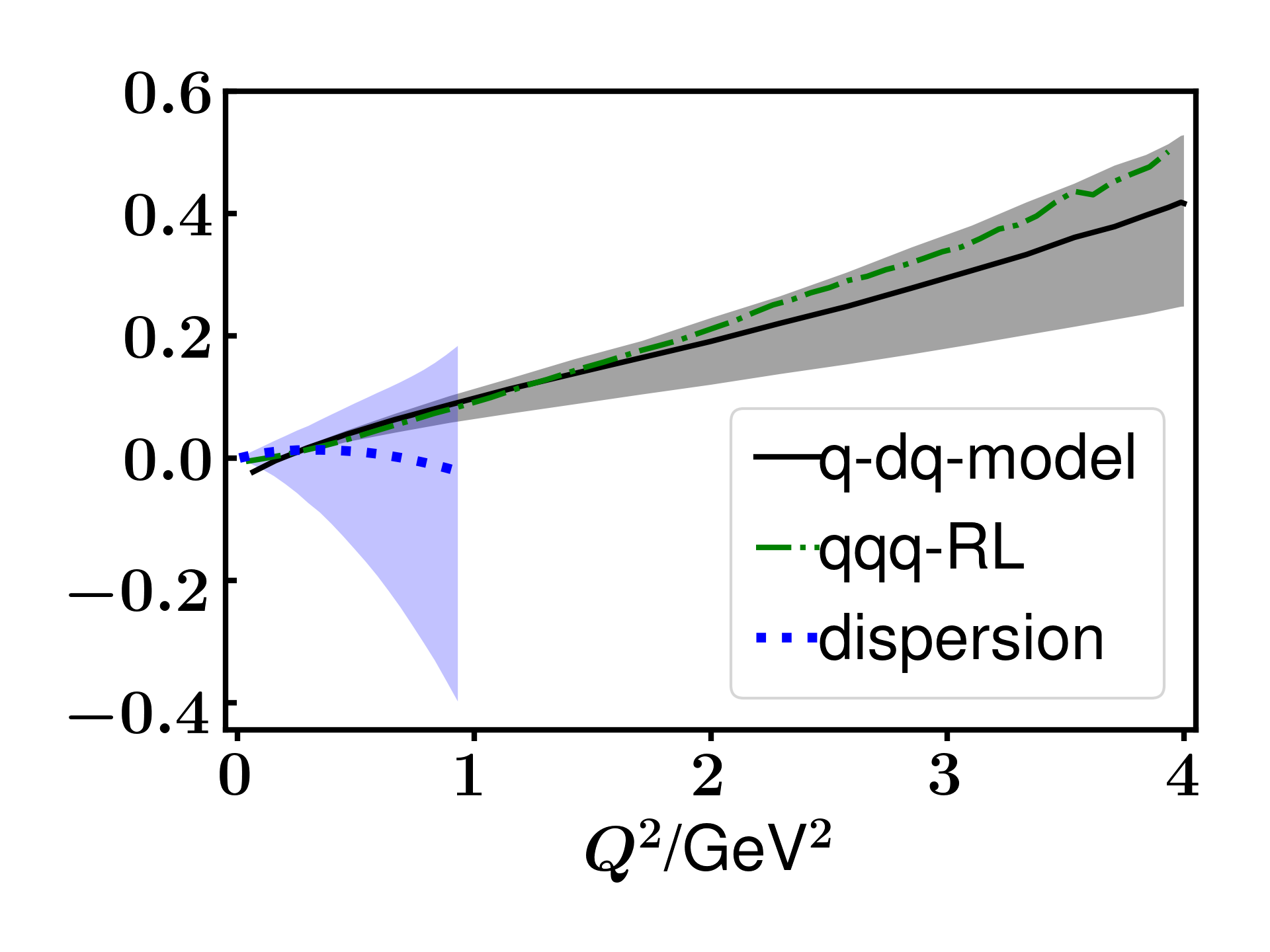}
		\end{center}
		\caption{The electromagnetic transition form factors $G_E$ (upper panel, left) and $G_M$ (upper panel, right) 
			and their ratio (lower panel) of 
			the process $\Sigma^0 \rightarrow \Lambda$. The magnetic form factor is again expressed in units of the 
			nuclear magneton. The black lines are our results from the quark-diquark Faddeev equation ('q-dq-model'). 
			The black bands	represent an error-estimate based on the variation of diquark mass parameters by 5\%. 
			The green dash-dot lines are the results from the three-body Faddeev equation ('qqq-RL') again using an effective
			running coupling in rainbow-ladder ('RL') approximation	discussed in Ref.~\cite{Sanchis-Alepuz2018b}. 
			The dispersive result (blue dotted line with blue error band) is taken from Ref.~\cite{Lin:2022dyu}.
			\label{fig:emff-sigma2lambda}}
	\end{figure}

	With the normalized Faddeev amplitudes of $\Sigma^0$ and $\Lambda$ at hand, we can go further and calculate 
	the electromagnetic transition form factors of $\gamma^{(*)} \Sigma^0 \rightarrow \Lambda$ in the space-like momentum region.
	Our results are shown in Fig.~\ref{fig:emff-sigma2lambda} and compared with corresponding results from the three-body
	Faddeev equation approach \cite{Sanchis-Alepuz2018b}. At non-vanishing momenta, these form factors have also been 
	studied in \cite{Ramalho:2012ad} with particular emphasis on disentangling contributions from the quark core with 
	meson cloud contributions, and in Ref.~\cite{Granados:2017cib} using dispersive methods. In Fig.~\ref{fig:emff-sigma2lambda}
	we also included latest results from the dispersive analysis of Ref.~\cite{Lin:2022dyu} with the value of $G_M$ at 
	zero momentum fixed in Ref.~\cite{Granados:2017cib}.
	 
	In principle, the electric transition form factor $G_E$ of $\gamma^{(*)} \Sigma^0 \rightarrow \Lambda$ should be exactly 
	equal to zero at the on-shell point of the photon $Q^2 = 0$, as can be seen from the dispersive results \cite{Lin:2022dyu}. 
	In practice, numerical errors are large at this point and consequently
	this is neither exactly the case in our calculation, nor in the three-body Faddeev results. Our numerical error
	is somewhat larger than in the three-body case but still tolerable. Overall, we obtained results for $G_E$ with
	the same qualitative behaviour as the one from the three-body Faddeev equation: the general magnitude of $G_E$ is small 
	and decrease slowly in the medium momentum region. Differences in magnitude in the medium and large momentum region are 
	again visible similar as for the elastic form factors. Overall we find from both approaches that the electric transition 
	of $\gamma^{(*)}\Sigma^0 \rightarrow \Lambda$ falls of slowly and therefore the transition is quite localized. 
	
	Note that we also further confirm an important difference to quark model calculations already discussed 
	in \cite{Sanchis-Alepuz2018b}: whereas in typical quark model setups the baryons are entirely composed of s- and d-waves,
	sizeable contributions from p-waves appear naturally in both, the relativistic three-body and the relativistic quark-diquark 
	Bethe-Salpeter approach. These are responsible for our non-zero result of the electric transition form factor. 
	In non-relativistic quark-models, similar results can only be obtained invoking unnaturally strong d-wave 
	contributions or including meson cloud effects \cite{Ramalho:2012ad}. 	
	
	As for the magnetic transition form factor $G_M$ of 
	$\gamma^{(*)}\Sigma^0 \rightarrow \Lambda$, the result calculated from the quark-diquark Faddeev equation is about a 
	factor of two smaller at vanishing momentum than the one calculated from the three-body Faddeev equation. The dispersive 
	value, estimated in Ref.~\cite{Granados:2017cib} from the experimental width of the decay $\Sigma^0 \rightarrow \Lambda \gamma$, 
	is of the same order of magnitude than the three-body result.\footnote{Note that Refs.~\cite{Granados:2017cib,Lin:2022dyu} use
		a different normalisation than we do, thus their $\kappa \approx 1.98$ corresponds to our 
		$\kappa \frac{2 m_p}{m_\Lambda+m_\Sigma} \approx 1.61$.} At finite momenta, the error of the dispersive result increases
	rapidly. Thus, given the current level or precision, dispersion theory cannot reasonably discriminate between our 
	quark-diquark model result and the three-body result. At medium and large momenta $Q^2 \geq 2$ GeV$^2$, the $G_M$ 
	calculated from quark-diquark Faddeev equation and the three-body Faddeev equation meet each other. In the third 
	diagram of Fig.~\ref{fig:emff-sigma2lambda} we also show the form factor ratio, which rises monotonically as momentum 
	increases. In this plot we did not include the error estimate of the dispersive result, but it is clear from the 
	upper panel that the dispersive result can only be trusted for very small momenta.

	\section{Summary}
	\label{sec:summary}

	In this work, we extended the quark-diquark Faddeev approach to the strange baryon sector and determined the elastic 
	electromagnetic form factors of $\Lambda(1116)$, $\Sigma^+(1189)$, $\Sigma^0(1193)$ and $\Sigma^-(1197)$. Our results 
	agree well on a quantitative level with lattice QCD calculations, where available. Interesting differences appear when 
	comparing our results with that calculated from a three-body Faddeev approach. We find the same overall running, albeit
	with sizeable differences in magnitude in certain momentum ranges that are different for different baryons. In general,
	the quark-diquark Faddeev method predicts harder form factors than that from three-body Faddeev approach calculations.
	
	A very interesting observation is the different running with photon momentum of the ratio of EMFFs for $\Sigma^{+,0}$ and $\Sigma^-$:	for $\Sigma^+$ and $\Sigma^0$ it decreases monotonically while for $\Sigma^-$ it remains constant. This can be 
	explained by the different diquark correlations inside these hyperons. 
	
	Furthermore, we computed the electromagnetic transition form factors for $\gamma^{(*)}\Sigma^0 \rightarrow \Lambda$. 
	The obtained results from quark-diquark Faddeev model have the same behavior as those calculated from the three-body 
	Faddeev approach, however again we observe sizeable difference in magnitude. Overall, we find a localized electric 
	transition form factor in both approaches.
	
	In the future, we plan to extend our calculations from the space-like to the time-like photon momentum. To this end we 
	also plan to include effects due to the appearance of vector meson resonances in the transverse part of the quark-photon 
	vertex that are currently not taken into account. It is furthermore useful to compare to other electromagnetic transition 
	form factors such as $\Delta \rightarrow N$ and $\Sigma(1385) \rightarrow \Lambda$ to obtain more insight into the role 
	played by the strange quark inside baryons.

	\acknowledgments{We are grateful for helpful discussions with G. Eichmann, P. Cheng and Z. Yao. 
	This work is supported by a Sino-German (CSC-DAAD) Postdoc Scholarship and project funding from the  
	Helmholtz Forschungsakademie Hessen für FAIR (HFHF).}

	\appendix

	\section{Computing ingredients for quark-diquark Faddeev equation}
	\label{sec:ingredients}
	
	\subsection{The basis for quark-diquark Faddeev amplitudes of $\Lambda$ and $\Sigma$}
	\label{subsec:basis}
	
	Denoting 
	\begin{equation}
		p_\perp^\mu = p^\mu - \hat{P}^\mu \angl{p \cdot \hat{P}},\, \gamma_\perp^\mu = \gamma^\mu - \hat{P}^\mu \angl{\gamma \cdot \hat{P}},\, \hat{P}^\mu = \dfrac{P^\mu}{i M}\,,
	\end{equation}
	then for the baryon with total spin $J = \frac{1}{2}$ and positive parity, the matrix valued functions for 
	scalar and axial-vector diquarks can be expanded as:
	\begin{subequations}
		\begin{align}
			&\Phi_{0^+}^{}(p, P) = \sum_{i=1}^{2} f_{0^+}^i(p, P) \mathcal{S}_i(p, P) \,, \\
			&\Phi_{1^+}^{\mu}(p, P) = \sum_{i=1}^{6} f_{1^+}^i(p, P) \mathcal{A}_i^\mu(p, P) \,,
		\end{align}
	\end{subequations}
	with dressing functions $f^i_{0^+,1^+}$ and Dirac tensors
	\begin{subequations}
		\begin{align}
			& \mathcal{S}_1(p, P) = \Lambda^+ \,,\\
			& \mathcal{S}_2(p, P) = i \gamma \cdot p^\perp \Lambda^+ \,, 
		\end{align}
	\end{subequations}
	\begin{subequations}
		\begin{align}
			&\mathcal{A}_{1\mu}(p, P) = -\gamma_5 \gamma^\perp_\mu \Lambda^+ \,, \\
			&\mathcal{A}_{2\mu}(p, P) = i P_\mu \gamma_5 \Lambda^+ \,, \\
			&\mathcal{A}_{3\mu}(p, P) = i p^\perp_\mu \gamma_5 \Lambda^+ \,, \\
			&\mathcal{A}_{4\mu}(p, P) = -i \gamma_5 \gamma^\perp_\mu \gamma\cdot p^\perp \Lambda^+ \,, \\
			&\mathcal{A}_{5\mu}(p, P) = -P_\mu \gamma_5 \gamma \cdot p^\perp \Lambda^+ \,, \\
			&\mathcal{A}_{6\mu}(p, P) = - p^\perp_\mu \gamma_5 \gamma \cdot p^\perp \Lambda^+ \,.
		\end{align}
	\end{subequations}
	
	\subsection{Diquark flavor wave function}
	\label{subsec:dq-flavor}
	
	We denoted the flavor wave functions of diquarks as $\lambda^a$: 
	\begin{align}
		&\lambda^{\sangl{ud}} = \dfrac{1}{\sqrt{2}}\begin{bmatrix}
			0 & 1 & 0 \\
			-1 & 0 & 0 \\
			0 & 0 & 0
		\end{bmatrix} \,, 
		\lambda^{\sangl{ds}} = \dfrac{1}{\sqrt{2}}\begin{bmatrix}
			0 & 0 & 0 \\
			0 & 0 & 1 \\
			0 & -1 & 0
		\end{bmatrix} \,, 
		\lambda^{\sangl{su}} = \dfrac{1}{\sqrt{2}}\begin{bmatrix}
			0 & 0 & -1 \\
			0 & 0 & 0 \\
			1 & 0 & 0
		\end{bmatrix} \,, \\
		&\lambda^{\bangl{uu}} = \begin{bmatrix}
			1 & 0 & 0 \\
			0 & 0 & 0 \\
			0 & 0 & 0
		\end{bmatrix} \,, 
		\lambda^{\bangl{ud}} = \dfrac{1}{\sqrt{2}}\begin{bmatrix}
			0 & 1 & 0 \\
			1 & 0 & 0 \\
			0 & 0 & 0
		\end{bmatrix} \,, 
		\lambda^{\bangl{dd}} = \begin{bmatrix}
			0 & 0 & 0 \\
			0 & 1 & 0 \\
			0 & 0 & 0
		\end{bmatrix} \,, \\
		&\lambda^{\bangl{su}} = \dfrac{1}{\sqrt{2}}\begin{bmatrix}
			0 & 0 & 1 \\
			0 & 0 & 0 \\
			1 & 0 & 0
		\end{bmatrix} \,, 
		\lambda^{\bangl{ds}} = \dfrac{1}{\sqrt{2}}\begin{bmatrix}
			0 & 0 & 0 \\
			0 & 0 & 1 \\
			0 & 1 & 0
		\end{bmatrix} \,, 
		\lambda^{\bangl{ss}} = \begin{bmatrix}
			0 & 0 & 0 \\
			0 & 0 & 0 \\
			0 & 0 & 1
		\end{bmatrix} \,, 
	\end{align}

	\section{Computing the electromagnetic form factors in quark-diquark Faddeev equation approach}
	\label{sec:emffs-ingredients}
	
	\begin{figure}[h]
		\centering
		\includegraphics[width=0.32\linewidth]{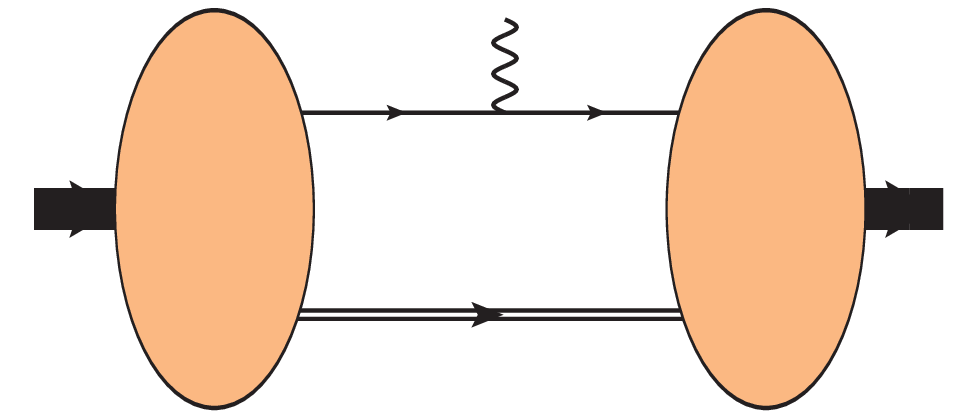}
		\includegraphics[width=0.32\linewidth]{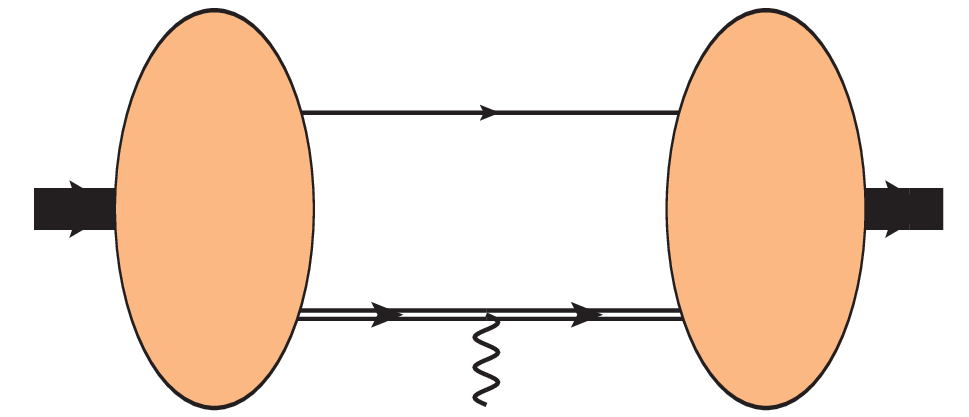}
		\\
		\includegraphics[width=0.32\linewidth]{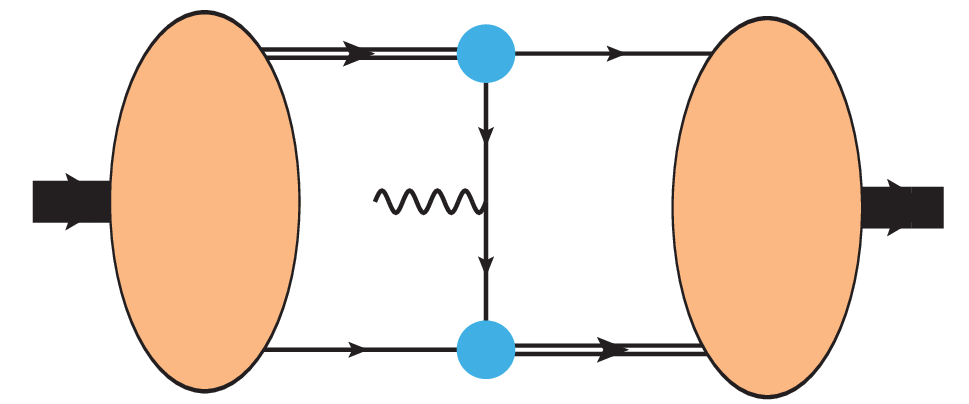} \hfill
		\includegraphics[width=0.32\linewidth]{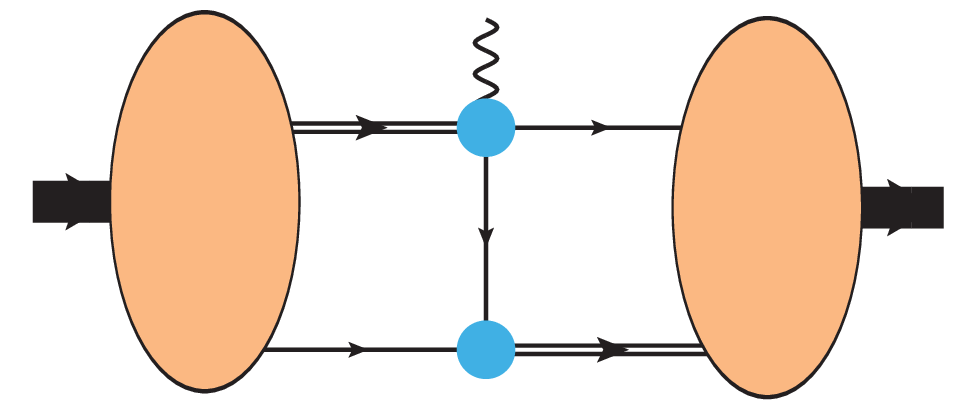} \hfill
		\includegraphics[width=0.32\linewidth]{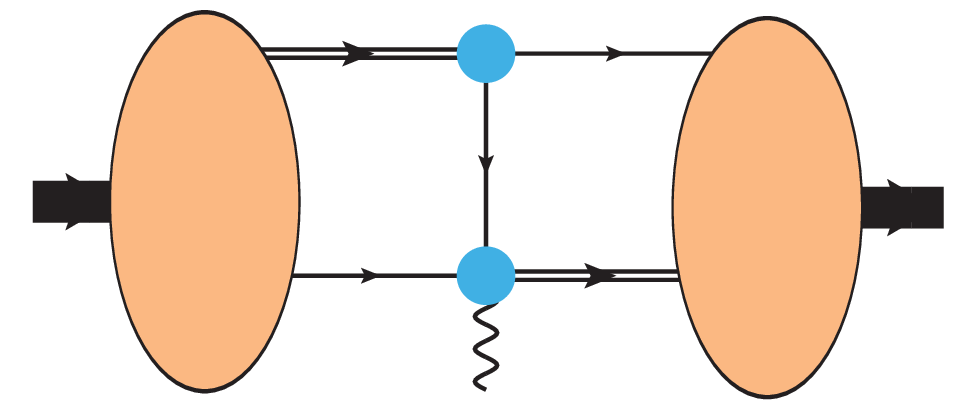}
		\caption{The diagrams of baryon electromagnetic current operator in the quark-diquark Faddeev equation.}
		\label{fig:nucleonelectromagneticcurrent}
	\end{figure}
	The electromagnetic current in the quark-diquark Faddeev equation has five types of diagrammatic contributions 
	arising from the gauging technique \cite{Haberzettl:1997jg,Kvinikhidze:1998xn,Kvinikhidze:1999xp,Oettel2000e,Oettel2000f,Eichmann2009b}. 
	The photon can interact: (1), with a bystander quark; (2), with a diquark (including 
	the photon induced transition between different diquarks); (3) with the exchanged quark; (4) with diquark the amplitude 
	and (5) with the conjugated diquark amplitude. These are depicted in Fig.~\ref{fig:nucleonelectromagneticcurrent}.
	Writing down the expressions explicitly, they are given by
	\begin{equation}
		J^\mu_{\alpha\beta}(Q^2) = \int_{p_f}\int_{p_i}\left[ \bar{\Phi}^a (p_f,P_f) X^{\mu\,, a b}(p_f,p_i,P_f,P_i) \Phi^b(p_i,P_i)\right] _{\alpha\beta}\,.
	\end{equation}
	Here $ p_i,p_f $ are the relative momentum of initial and final states and $ P_i,P_f $ are the total momentum of initial and final states, $ P_f = P_i + Q $. $ \alpha, \beta = 1,\cdots, 4 $ are the quark indices and $ a,b $ are the diquark indices. 
	The quantity $ X^{\mu\,, a b}(p_f,p_i,P_f,P_i) $ is given by
	\begin{equation}
		\begin{split}
			X^{\mu\,, a b} &= X^{\mu\,, a b}_q \; (2\pi)^4 \delta^4(p_f - p_i - (1-\eta)Q) \\
			&\quad + X^{\mu\,, a b}_{dq}\; (2\pi)^4 \delta^4(p_f - p_i + \eta Q) + X^{\mu\,, a b}_{K}\,.
		\end{split}
	\end{equation}
	\begin{align}
		&X^{\mu\,, a b}_{q,\,\alpha\beta} = \left[ S(p_+)\Gamma^\mu_q(p_+,p_-)S(p_-)\right]_{\alpha\beta} D^{ab}(p_{d-})\,,\\
		&X^{\mu\,, a b}_{dq,\,\alpha\beta} = [S(p_-)]_{\alpha\beta} \left[ D^{aa'}(p_{d+})\Gamma_{dq}^{\mu,\,a'b'}D^{b'b}(p_{d-})\right] \,,\\
		&X^{\mu\,, a b}_{K,\,\alpha\beta} = D^{aa'}(p_{d+}) \left[ S(p_+) K^{\mu,\,a'b'} S(p_-) \right] _{\alpha\beta} D^{b'b}(p_{d-})\,,
	\end{align}
	where
	\begin{align*}
		&p_- = p_i + \eta P_i\,,\qquad p_{d-} = -p_i + (1-\eta)P_i\,,\\
		&p_+ = p_f + \eta P_f\,, \qquad p_{d+} = -p_f + (1-\eta) P_f\,.
	\end{align*}
	As for the gauged kernel,
	\begin{equation}
		K^{\mu,\,ab} = K^{\mu,\,ab}_{EX} + K^{\mu,\,ab}_{SG} + K^{\mu,\,ab}_{\overline{SG}}\,,
	\end{equation}
	where
	\begin{align}
		&K^{\mu,\,ab}_{EX} = \Gamma^b_{dq}(p_1,p_{d-}) \left[ S(q')\Gamma_q^\mu(q',q)S(q)\right] ^T \bar{\Gamma}^a(p_2,p_{d_+})\,,\\
		&K^{\mu,\,ab}_{SG} = M^{\mu,\,b}(k_1,p_{d-},Q)S^T(q') \bar{\Gamma}^a(p_2,p_{d+})\,,\\
		&K^{\mu,\,ab}_{\overline{SG}} = \Gamma^b(p_1,p_{d-}) S^T(q) \bar{M}^{\mu,\,a}(k_2,p_{d+},Q)\,,
	\end{align}
	and
	\begin{align*}
		&q = p_{d-} - p_+\,,\qquad p_1 = \dfrac{p_{+}-q}{2}\,,\qquad p_2 = \dfrac{p_- - q'}{2}\,,\\
		&q' = p_{d+} - p_-\,,\qquad k_1 = \dfrac{p_+ - q'}{2}\,,\qquad k_2 = \dfrac{p_- - q}{2}\,.
	\end{align*}
	In the following we discuss these diagrams in turn. 
	
	\subsection{Quark-photon diagram}

	The quark-photon current contribution is
	\begin{equation}
		\begin{split}
			J^\mu_{1}(Q^2) 
			&= \int_{p_i} \left[ \bar{\Phi}^a (p_f,P_f) \left[ S(p_+)\Gamma^\mu_q(p_+,p_-)S(p_-)\right] D^{ab}(p_{d-}) \Phi^b(p_i,P_i)\right]\,,
		\end{split}
	\end{equation}
	with $ p_f = p_i + (1-\eta)Q $. The quark-photon vertex involved in this diagram has to satisfy the Ward-Takahashi identity:
	\begin{equation}
		i Q_\mu \Gamma^\mu(p_2,p_1) = S^{-1}(p_2) - S^{-1}(p_1)\,.
	\end{equation}
	To this end we employ the Ball-Chiu solution of this identity:
	\begin{equation}
		\Gamma^\mu(p_2,p_1) = \gamma^\mu \Sigma_A + (p_1+p_2)^\mu \left[ \dfrac{\slashed p_1 + \slashed p_2}{2}\triangle_A - i\triangle_B\right] + \Gamma^\nu_T(p2,p1) \,,
	\end{equation}
	with
	\begin{equation}\label{delta-def}
		\Sigma_F(p_2,p_1) = \dfrac{F(p_2^2)+F(p_1^2)}{2}\,,\qquad \triangle_F = \dfrac{F(p_2^2)-F(p_1^2)}{p_2^2-p_1^2}\,.
	\end{equation}
	The eight amplitudes in the transverse parts $\Gamma^\nu_T(p2,p1)$ cannot be constrained by the WTI and we ignore 
	them in this work. Note, however, that due to the appearance of vector meson contributions, these will certainly 
	play an important role in the time-like momentum region and we therefore plan to include those in future work.
	
	\subsection{Diquark-photon diagram}
	\label{sec:diquark-photon}
	
	The diquark-photon current contribution is
	\begin{equation}
		\begin{split}
			J^\mu_{\alpha\beta}(Q^2) 
			&= \int_{p_i} \left[ \bar{\Phi}^a (p_f,P_f) [S(p_-)]_{\alpha\beta} \left[ D^{aa'}(p_{d+})\Gamma_{dq}^{\mu,\,a'b'}D^{b'b}(p_{d-})\right] \Phi^b(p_i,P_i)\right]\,,
		\end{split}
	\end{equation}
	with $ p_f = p_i - \eta Q $. The diquark-photon vertex again satisfies a Ward-Takahashi identity.
	For the scalar diquark, it is the electromagnetic vertex for spin-0 particle:
	\begin{equation}
		Q_\mu \Gamma^\mu_{sc}(k,Q) = D^{-1}(k_+) - D^{-1}(k_-) = Q_\mu \left[ 2 k^\mu \triangle_{D^{-1}}\right] \,,
	\end{equation}
	with $ Q = k_+ - k_- $ and $ k = (k_- + k_+) / 2 $ and $\triangle_{D^{-1}}$ is defined as in Eq.~(\ref{delta-def}). 
	This relation is strictly valid for point-like particles. Considering the internal structure of our scalar diquarks, 
	we added the following form factor:
	\begin{equation}
		\Gamma^\mu_{0^+}(k,Q) = 2k^\mu \triangle_{D_{0^+}^{-1}} \dfrac{1}{1 + Q^2 / \Lambda_{0^+}^2}\,,
	\end{equation}
    The scale $\Lambda_{0^+}^2$ is a free parameter and needs to be determined by comparison with experimental data for the
    electromagnetic form factors. Here, we choose the electromagnetic form factors of nucleons as benchmarks and obtain 
    $\Lambda_{0^+}^2 = 1.024$ GeV$^2$. 
	
	For axial vector diquark, we adopted the form of electromagnetic vertex for spin-1 particle \cite{Bhagwat2008a}.
	The axial vector diquark-photon vertex is (denote $T_{\mu\nu}(Q) = \delta_{\mu\nu} - Q_\mu Q_\nu / Q^2$)
	\begin{align}
		&\Gamma^{\mu,\,\alpha\beta}_{1^+}(k,Q) = (p_i + p_f)^\mu T^{\alpha\gamma}(p_f) T^{\gamma\beta}(p_i) \triangle_{D_{1^+}^{-1}} F_1(Q^2) \\
		&\nonumber\qquad\qquad\qquad + \left( T^{\mu\beta}(p_i) T^{\alpha\gamma}Q_\gamma - T_{\mu\alpha}(p_f)T_{\beta\gamma}Q_\gamma \right) \triangle_{D_{1^+}^{-1}} F_2(Q^2) \\
		&\nonumber\qquad\qquad\qquad - (p_i + p_f)^\mu T^{\alpha\gamma}(p_f)Q_\gamma T^{\beta\lambda}(p_i)Q_\lambda \triangle_{D_{1^+}^{-1}} F_3(Q^2) \,.
	\end{align}
	The $F_i(Q^2)$ relate with the multipole form factors of axial vector diquark as
	\begin{align}
		&F_1(Q^2) = G_e(Q^2) - \dfrac{2}{3} \tau G_q(Q^2) \,, \tau = \dfrac{Q^2}{4M_{1^+}^2}\,,\\
		&F_2(Q^2) = G_m(Q^2) \,, \\
		&F_3(Q^2) = \dfrac{1}{2 M_{1^+}^2}\dfrac{-G_e(Q^2) + G_m(Q^2) + (1 + 2/3 \tau) G_q(Q^2)}{1 + \tau} \,.
	\end{align}
	We parameterized the multipole form factors of axial vector diquark as
	\begin{align}
		&G_e(Q^2) = \dfrac{1}{\left( 1 + Q^2 / \Lambda_{1^+}^2\right) ^2} \,, G_m(Q^2) = \dfrac{g_m}{\left( 1 + Q^2 / \Lambda_{1^+}^2\right) ^2} \,, G_q(Q^2) = \dfrac{-g_q}{\left( 1 + Q^2 / \Lambda_{1^+}^2\right) ^2} \,.
	\end{align}
	Similar as above, we determine the free parameter by best fits to the electromagnetic form factors of nucleons. 
	We obtain $\Lambda_{1^+}^2 = 0.582$ GeV$^2$, $g_m = 1.21$ and $g_q = 0.899$. 
	
	The last diquark-photon vertex is the scalar-axial vector diquark transition vertex, which cannot be obtained from gauged technique. This vertex is purely transverse and cannot be constrained by the current conservation:
	\begin{equation}\label{}
		Q_\mu \Gamma^{\mu,\,5 \beta}_{sa} = Q_\mu \Gamma^{\mu,\,\alpha 5}_{as} = 0\,.
	\end{equation}
	In analogy to the $ \rho \rightarrow \pi \gamma $ decay, the transition vertex is given by the Lorentz structure
	\begin{equation}\label{}
		\Gamma^{\mu,\,5 \beta}_{sa}(p_2,p_1) = i \varepsilon^{\mu\beta\rho\sigma} Q^\rho (p_1 + p_2)^\sigma \dfrac{\kappa_{sa}}{2 M_{1^+}}\dfrac{1}{\left( 1 + Q^2 / \Lambda_{sa}^2\right) ^2}\,,
	\end{equation}
	\begin{equation}\label{}
		\Gamma^{\mu,\,\alpha 5}_{as}(p_2,p_1) = -i \varepsilon^{\mu\beta\rho\sigma} Q^\rho (p_1 + p_2)^\sigma \dfrac{\kappa_{sa}}{2 M_{1^+}}\dfrac{1}{\left( 1 + Q^2 / \Lambda_{sa}^2\right) ^2}\,.
	\end{equation}
	Again, best fit results for the proton and neutron electromagnetic form factor are obtained 
	for $\Lambda_{sa}^2 = 0.871$ GeV$^2$ and $\kappa_{sa} = 0.470$.
	
	\subsection{Exchange quark-photon diagram}

	The explicit expression for the exchange quark-photon diagram contribution is given by 
	\begin{equation}
		\begin{split}
			J^\mu_{\alpha\beta}(Q^2) &= \int_{p_f}\int_{p_i}\left[ \bar{\Phi}^a (p_f,P_f) D^{aa'}(p_{d+})\right.\\
			&\quad\quad\qquad\left. \left[ S(p_+) \Gamma^{b'}_{dq}(p_1,p_{d-}) \left[ S(q')\Gamma_q^\mu(q',q)S(q)\right] ^T \bar{\Gamma}^{a'}(p_2,p_{d_+}) S(p_-) \right] _{\alpha\beta} \right.\\
			&\left.\quad\quad\qquad D^{b'b}(p_{d-}) \Phi^b(p_i,P_i)\right]\,.
		\end{split}
	\end{equation}
	All ingredients for this diagram have been discussed already above.

	\subsection{Seagull diagram}
	
	\begin{figure}[t]
		\includegraphics[width=0.32\linewidth]{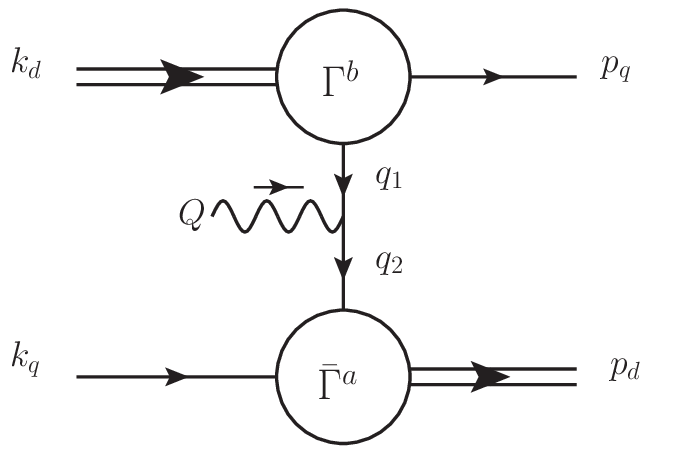} \hfill
		\includegraphics[width=0.32\linewidth]{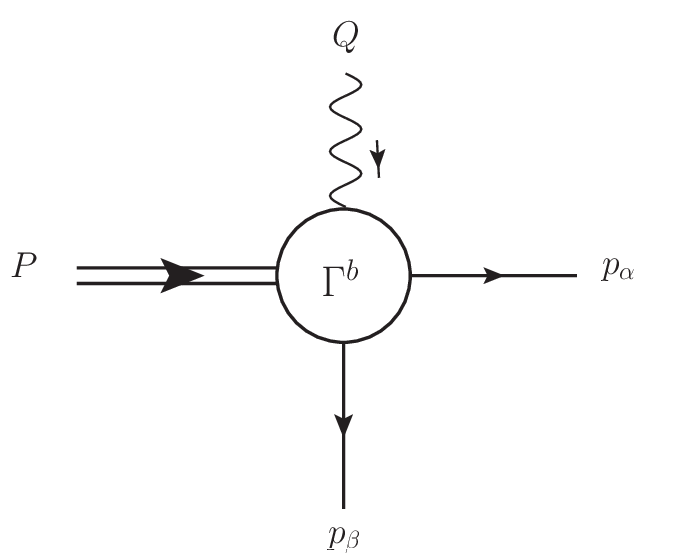} \hfill
		\includegraphics[width=0.32\linewidth]{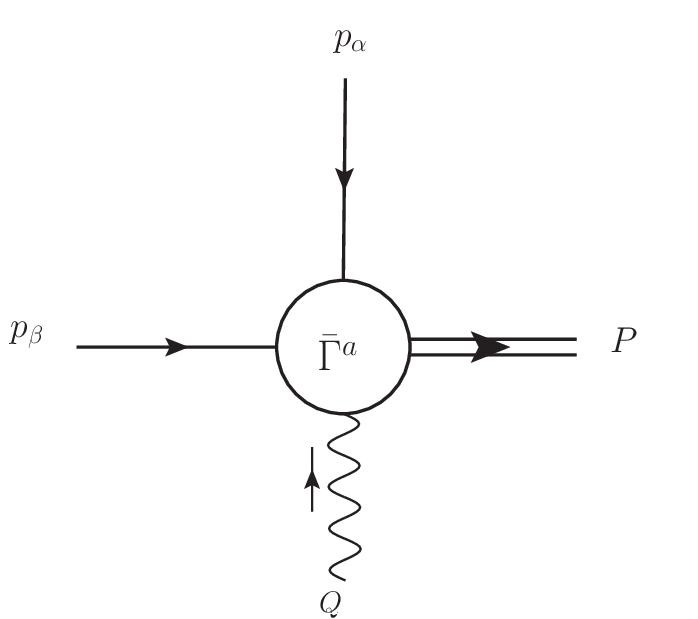}
		\caption{The Feynman diagrams of exchange (left), seagull (center) and seagullbar (right) in the quark-diquark Faddeev equation. \label{fig:ex-seg}}
	\end{figure}
	The diquark amplitude-photon diagram contribution is given explicitly by (the corresponding Feynman diagrams have been
	shown in Fig.~\ref{fig:nucleonelectromagneticcurrent} and are partially repeated in more detail in Fig.~\ref{fig:ex-seg})
	\begin{equation}
		\begin{split}
			J^\mu_{\alpha\beta}(Q^2) &= \int_{p_f}\int_{p_i}\left[ \bar{\Phi}^a (p_f,P_f) D^{aa'}(p_{d+})\right.\\
			&\quad\quad\qquad\left. \left[ S(p_+) M^{\mu,\,b'}(k_1,p_{d-},Q)S^T(q') \bar{\Gamma}^{a'}(p_2,p_{d+}) S(p_-) \right] _{\alpha\beta} \right.\\
			&\left.\quad\quad\qquad D^{b'b}(p_{d-}) \Phi^b(p_i,P_i)\right]\,.
		\end{split}
	\end{equation}
	Here we need consider the vertex of a photon interacting with the diquark amplitudes. Again, these vertices have to
	satisfy a Ward-Takahashi identity and we rely again on a Ball-Chiu construction. To this end we only 
	consider the leading term of diquark amplitudes, Eq.~\eqref{eq:dq-amp}. The gauged quark-diquark vertex is 
	then given by (here we omit the flavor wave functions)
	\begin{align}
		& M^\mu(p,P,Q) = \left[ e_- (p+Q/4)^\mu \triangle_{\mathcal{F}}^+ - e_+ (p-Q/4)^\mu \triangle_{\mathcal{F}}^-\right] i \gamma_5 C\,,\\
		& M^{\mu, \alpha}(p,P,Q) = \left[ e_- (p+Q/4)^\mu \triangle_{\mathcal{F}}^+ - e_+ (p-Q/4)^\mu \triangle_{\mathcal{F}}^-\right] i\gamma^\alpha C\,,
	\end{align}
	where
	\begin{equation}
		\triangle_{\mathcal{F}}^\pm = \dfrac{\mathcal{F}(p_\pm^2)-\mathcal{F}(p^2)}{p_\pm^2 - p^2} = \pm \dfrac{\mathcal{F}(p_\pm^2)-\mathcal{F}(p^2)}{Q\cdot(q\pm Q/4)}\,,
	\end{equation}
	and $e_-, e_+$ is the charge-flavor factor computed from flavor wave functions. For details, see e.g. Ref.~\cite{Eichmann2009b}.

	\subsection{Seagull bar diagram}
	
	The conjugated diquark amplitude-photon diagram contribution is given by (the sketched Feynman diagram is shown in
	the right diagram of Fig.~\ref{fig:ex-seg})
	\begin{equation}
		\begin{split}
			J^\mu_{\alpha\beta}(Q^2) &= \int_{p_f}\int_{p_i}\left[ \bar{\Phi}^a (p_f,P_f) D^{aa'}(p_{d+})\right.\\
			&\quad\quad\qquad\left. \left[ S(p_+) \Gamma^{b'}(p_1,p_{d-}) S^T(q) \bar{M}^{\mu,\,a'}(k_2,p_{d+},Q) S(p_-) \right] _{\alpha\beta} \right.\\
			&\left.\quad\quad\qquad D^{b'b}(p_{d-}) \Phi^b(p_i,P_i)\right]\,.
		\end{split}
	\end{equation}
	Similar to the diquark amplitude-photon vertex,	the gauged conjugated diquark-quark vertex satisfies the following 
	relation with the gauged quark-diquark vertex
	\begin{align}
		\bar{M}^\mu(q,P,Q) = - C (M^\mu)^T (-q, -P, Q) C^{-1}\,,\\
		\bar{M}^{\mu,\,\alpha}(q,P,Q) = C (M^{\mu,\,\alpha})^T (-q, -P, Q) C^{-1}\,.
	\end{align}
	They are given by
	\begin{align}
		& \bar{M}^\mu(p,P,Q) = \left[ e_- (p-Q/4)^\mu \triangle_f^- - e_+ (p+Q/4)^\mu \triangle_f^+\right] i \gamma_5 C\,,\\
		& \bar{M}^{\mu, \alpha}(p,P,Q) = \left[ e_- (p-Q/4)^\mu \triangle_f^- - e_+ (p+Q/4)^\mu \triangle_f^+\right] i \gamma^\alpha C\,.
	\end{align}
	
	\subsection{Results for the nucleon electromagnetic form factors}\label{nucleon}
	
	\begin{figure}[t]
	\centering
	\includegraphics[width=0.48\linewidth]{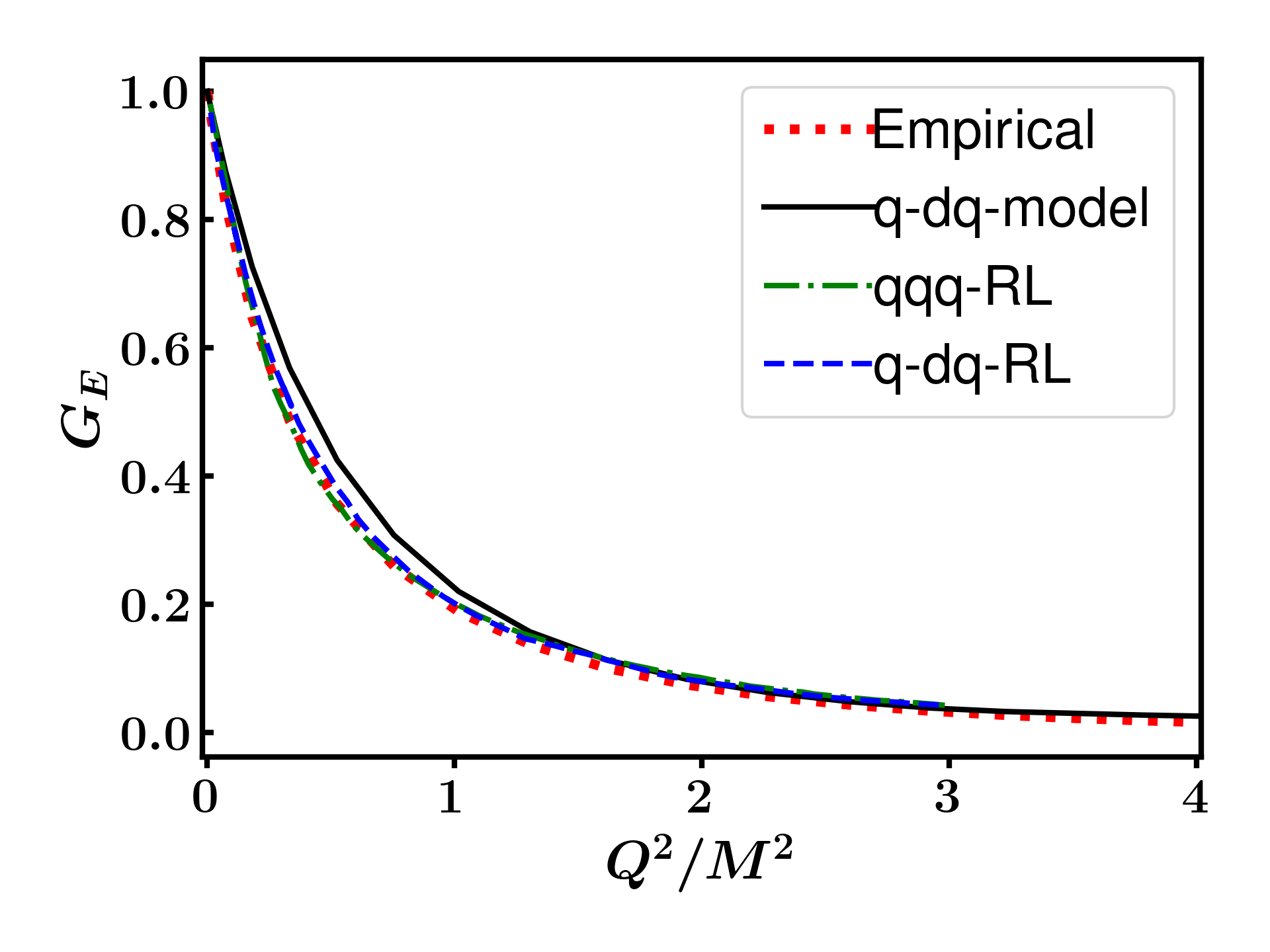} \hfill
	\includegraphics[width=0.48\linewidth]{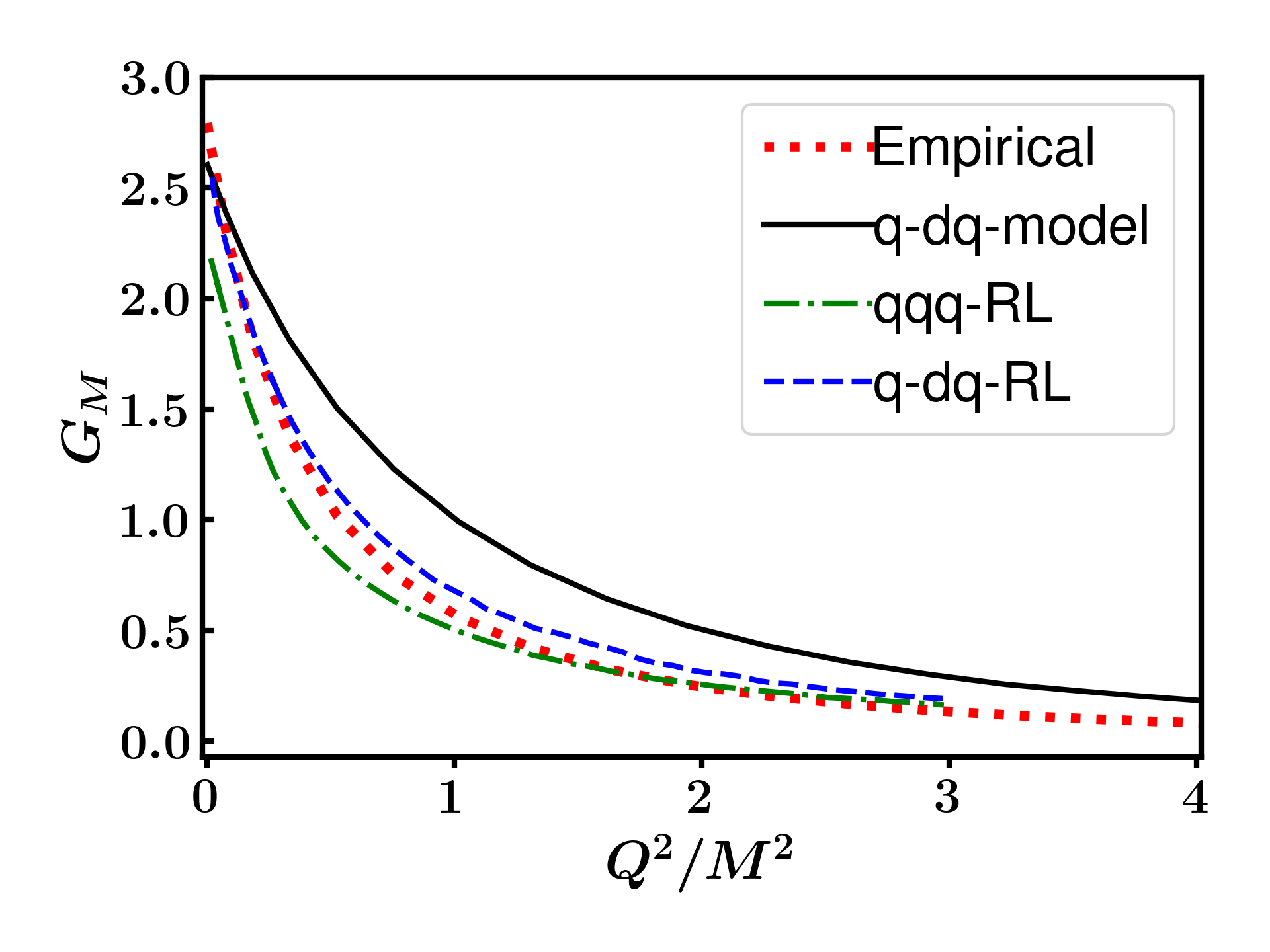} 
	\\
	\includegraphics[width=0.48\linewidth]{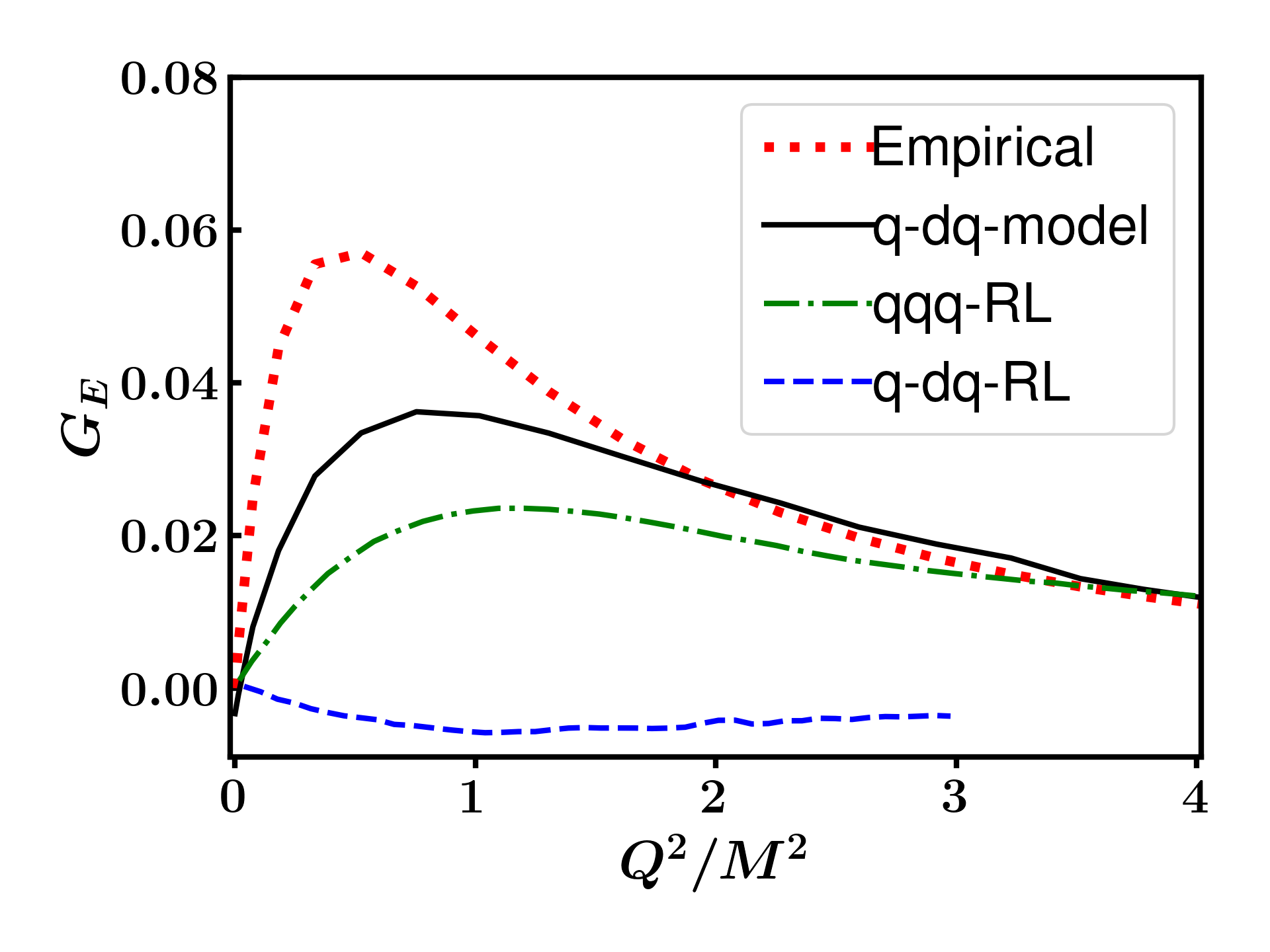} \hfill
	\includegraphics[width=0.48\linewidth]{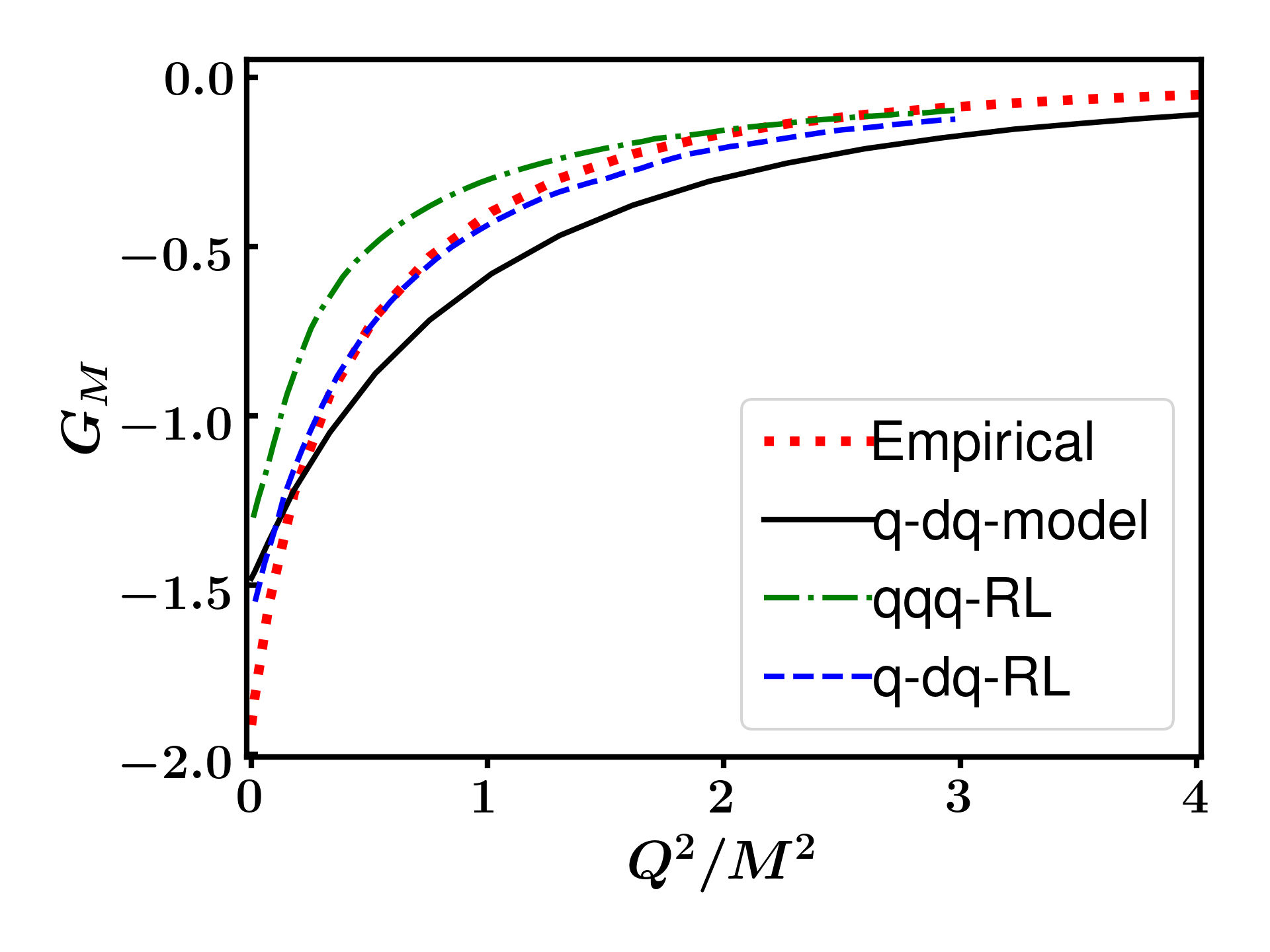}
	\caption{The comparison of the electric form factor $G_E$ (left) and magnetic form factor $G_M$ (right) of proton (top) 
	and neutron (bottom) which are calculated from quark-diquark Faddeev approach in this work (black solid lines, 'q-dq-model'), the
	three-body Faddeev approach (green dash-dotted lines, 'qqq-RL') using an effective running coupling in rainbow-ladder ('RL')
	approximation \cite{Eichmann2011b} and the quark-diquark Faddeev approach using the same effective running coupling (blue  
	lines, 'q-dq-RL') \cite{Eichmann2009b}. The empirical data (red dotted line) are adapted from \cite{Kelly:2004hm}.
	\label{fig:comp-emff-nucl} }
    \end{figure}

	In the previous subsections, we summarised the (well-known) framework for the calculation of electromagnetic form factors
	in the quark-diquark approach. In this framework parameters appear, which have to be fitted to experimental input. To this
	end we first calculated the nucleon electromagnetic form factors and determined values for the parameters by a best fit 
	to the available experimental data. 
	For simplicity, we selected the empirical fit from experimental data in Ref.~\cite{Kelly:2004hm} as basis for our experimental 
	input\footnote{We explicitly checked that fits to more modern representations of the data such as the ones given in 
	Ref.~\cite{Ye:2017gyb} do not change our results materially. We expect this to be also the case for the parametrisation given 
	in the very recent work Ref.~\cite{Lin:2021xrc}.}. 
	Our objective was to fit the data within the photon momentum range of $Q^2 \sim [1, 2] \,\text{GeV}^2$ in order to determine 
	the most optimal parameters. This specific range was chosen for two reasons. First, Due to the absence of explicit meson 
	cloud effects in the current formulation of the quark-diquark Faddeev framework, we cannot expect to perfectly describe 
	nucleons, particularly neutrons, in the infrared momentum range. Second, in the ultraviolet momentum range perturbative logarithmic running sets in, which has never been implemented in the quark-diquark model for simplicity. Hence, our 
	quark-diquark Faddeev approach is aptly suited for intermediate momentum range physics, which is why we specifically opted 
	for $Q^2 \sim [1, 2] \,\text{GeV}^2$.  
	
	Our results for the electric and magnetic form factors of the proton and the neutron are shown in Fig.~\ref{fig:comp-emff-nucl}
	('q-dq-model') together with results from a quark-diquark Faddeev approach using an effective quark-gluon interaction in rainbow-ladder approximation ('q-dq-RL') \cite{Eichmann2009b}, results from the three-body Faddeev approach using the same effective interaction 
	('qqq-RL') \cite{Eichmann2011b} and empirical fitted results from \cite{Kelly:2004hm}. The spread in the different DSE/BSE 
	approaches reflect overall systematic 
	uncertainties which are most prominent in the neutron electric form factors but also significant in all other form factors.
	These uncertainties at small momenta are related to the omission of meson cloud effects. These effects are non-universal and 
	may have a different quantitative impact in different truncation schemes. Their systematic inclusion in the	DSE/BSE framework 
	has been explored in a series of works both at zero and finite temperature, see e.g. \cite{Fischer:2007ze,Fischer:2008wy,Sanchis-Alepuz:2014wea,Gunkel:2021oya} and references therein. 
	
	In general we observe that for magnetic form factors the decreasing rates are always slower in the quark-diquark picture 
	than in the three-body picture in both proton and neutron cases. This may be due a clustering effect in diquarks which 
	reduces	the 'magnetic size' of the nucleon. In contrast to the quark-diquark model discussed in Ref.~\cite{Cloet2009b} we
	have optimized our parameters to best possible match the electric form factor of the neutron rather than the magnetic form
	factors, as can be seen from the lower panel of Fig.~\ref{fig:comp-emff-nucl}. On the flip side we have to live with 
	somewhat too hard magnetic form factors as compared to the empirical curve and the other BSE-approaches. This has
	to be kept in mind, when predicting form factors of strange baryons and is discussed again in the main par of this work.

	\section{Fitting parameters of the electromagnetic form factors}\label{app:fits}
	 
	 In Tab.~\ref{tab:fit-params} we provide results for fitting our numerical results for the electromagnetic form factors
	 of $\Lambda$ and $\Sigma$ with the fit function 
	\begin{equation}\label{eq:fit-form}
		G(Q^2) = \dfrac{n_0 + n_1 Q^2}{1 + d_1 Q^2 + d_2 Q^4 + d_3 Q^6}\,.
	\end{equation}
	in analogy to what has been done in Ref.~\cite{Sanchis-Alepuz2016a}.  
	
	\begin{table}[t]
		\centering
		\begin{tabular}{c|c|ccc|ccc}
			\hline
			\multicolumn{2}{c|}{} & \multicolumn{3}{c|}{$G_E$} & \multicolumn{3}{c}{$G_M$} \\
			\cline{3-8}
			\multicolumn{2}{c|}{}  & 95\%$m_{dq}$ & $m_{dq}$ & 105\%$m_{dq}$ & 95\%$m_{dq}$ & $m_{dq}$ & 105\%$m_{dq}$ \\
			\hline
			\multirow{5}{*}{$\Lambda$} & $n_0$ & 0 & 0 & 0 & -0.393 & -0.388 & -0.393 \\
			& $n_1$ & -0.158 & -0.167 & -0.229 & -0.717 & -0.667 & -0.451 \\
			& $d_1$ & 0.944 & 1.002 & 1.831 & 2.544 & 2.373 & 1.722 \\
			& $d_2$ & 0.578 & 0.525 & 0.397 & 0.444 & 0.412 & 0.288 \\
			& $d_3$ & 0.285 & 0.309 & 0.480 & 0.224 & 0.202 & 0.135 \\
			\hline
			\multirow{5}{*}{$\Sigma^+$} & $n_0$ & 1 & 1 & 1 & 2.602 & 2.465 & 2.243 \\
			& $n_1$ & -0.238 & -0.244 & -0.243 & 4.930 & 19.185 & 69.491 \\
			& $d_1$ & 1.690 & 1.800 & 1.916 & 3.300 & 9.386 & 33.030 \\
			& $d_2$ & 0.921 & 0.929 & 0.942 & 2.774 & 10.384 & 37.326 \\
			& $d_3$ & 0.195 & 0.134 & 0.108 & 0.868 & 3.359 & 12.808 \\
			\hline
			\multirow{5}{*}{$\Sigma^0$} & $n_0$ & 0 & 0 & 0 & 0.678 & 0.658 & 0.582 \\
			& $n_1$ & -0.201 & -0.236 & -0.272 & 0.059 & 10.071 & 47.353 \\
			& $d_1$ & 0.916 & 1.297 & 1.675 & 1.299 & 17.416 & 80.891 \\
			& $d_2$ & 0.972 & 1.019 & 1.048 & 0.385 & 16.899 & 86.352 \\
			& $d_3$ & 0.094 & 0.148 & 0.214 & 0.044 & 5.437 & 23.572 \\
			\hline
			\multirow{5}{*}{$\Sigma^-$} & $n_0$ & -1 & -1 & -1 & -1.250 & -1.159 & -1.039 \\
			& $n_1$ & -1.612 & -0.348 & -0.245 & -4.150 & -0.722 & -36.272 \\
			& $d_1$ & 3.111 & 1.894 & 1.865 & 4.913 & 2.187 & 37.664 \\
			& $d_2$ & 2.705 & 0.897 & 0.701 & 5.150 & 1.343 & 47.223 \\
			& $d_3$ & 0.981 & 0.265 & 0.223 & 2.096 & 0.429 & 21.592 \\
			\hline
		\end{tabular}
		\caption{The best fitting parameters of the elastic electromagnetic form factors of $\Lambda$- and $\Sigma$- baryon states using Eq.\eqref{eq:fit-form}. Here $m_{dq}$ means the mass parameter of diquarks. \label{tab:fit-params}}
	\end{table}
	
	\newpage
	
	\addcontentsline{toc}{section}{References}
	\bibliographystyle{unsrt}
	\bibliography{lambda-sigma0-emff} 

\end{document}